\documentclass[10pt]{article}
\usepackage{amsmath}
\usepackage{physics}
\usepackage{amsfonts}
\usepackage{amssymb}
\usepackage{graphicx}%
\usepackage{amstext}
\usepackage{latexsym}
\usepackage{color}
\usepackage{mathtools}
\usepackage{graphics}
\usepackage{cite}
\usepackage{slashed}
\usepackage{epsfig}
\usepackage{amsthm}
\usepackage{appendix}
\usepackage{hyperref}
\usepackage{multirow}
\usepackage{color}
\usepackage[normal]{subfigure}
\usepackage{rotating}
\usepackage[table]{xcolor}
\usepackage{enumitem}
\usepackage[utf8x]{inputenc}
\usepackage{authblk}
\usepackage{colortbl}
\usepackage{pdflscape}
\usepackage{bm}
\usepackage{gensymb}
\usepackage[english]{babel}

\allowdisplaybreaks[1]

\newcommand{\be}{\begin{equation}}
\newcommand{\en}{\end{equation}}

\newtheorem{defi}{Definition}[section]
\newtheorem{lem}[defi]{Lemma}

\newcommand{\bedefin}{\begin{defi}}
\newcommand{\findefi}{\end{defi} \medskip}

\newcommand{\belem}{\begin{lem}$\!\!${\bf }}
\newcommand{\enlem}{\end{lem}}

\newcommand{\beno}{\begin{equation*}}
\newcommand{\enno}{\end{equation*}}

\newcommand{\bea}{\begin{eqnarray}}
\newcommand{\ena}{\end{eqnarray}}

\catcode `\@=11 \@addtoreset{equation}{section}

\catcode `\@=12

\newcommand{\g}{G_{\hbox{\tiny{NC}}}}
\newcommand{\gd}{\hat{G}_{\hbox{\tiny{NC}}}}
\newcommand{\G}{\mathfrak{g}_{\hbox{\tiny{NC}}}}
\newcommand{\wh}{G_{\hbox{\tiny{WH}}}}
\newcommand{\WH}{\mathfrak{g}_{\hbox{\tiny{WH}}}}

\begin{document}

\title{Gauge invariant energy spectra in 2-dimensional noncommutative quantum mechanics}

\author[1,2]{S. Hasibul Hassan Chowdhury\thanks{shhchowdhury@bracu.ac.bd}}
\author[1,3]{Talal Ahmed Chowdhury\thanks{talal@du.ac.bd}}
\author[1,4]{Md Arman Ud Duha\thanks{arman.duha@okstate.edu}}

\affil[1]{Department of Physics, University of Dhaka, P.O. Box 1000, Dhaka, Bangladesh}
\affil[2]{Department of Mathematics and Natural Sciences, BRAC University, 66 Mohakhali, Dhaka 1212, Bangladesh}
\affil[3]{The Abdus Salam International Centre for Theoretical Physics, Strada Costiera 11, I-34151 Trieste, Italy}
\affil[4]{Department of Physics, Oklahoma State University, Stillwater, OK 74078, USA}
\date{\today}

\maketitle

\begin{abstract}
In this paper, we consider an electron moving on a 2D noncommutative plane immersed in a constant magnetic field under the influence of a polynomial potential. We obtain gauge invariant energy spectra of this system using 2-parameter family of unitarily equivalent irreducible representations of the nilpotent Lie group $\g$ that were worked out in detail in \cite{nctori}. We work out the cases of anisotropic harmonic potential and the Hall potential as physical applications of the proposed method. We also show subsequently that straightforward generalization of the Landau problem and the quantum Hall effect in the noncommutative setting using minimal coupling prescription as is done naively on many occasions in the literature violate gauge invariance of energy spectra.
\end{abstract}

\section{Introduction}\label{sec:intro}

The basic constituent of our study is a noncommutative phase space which is coordinatised by an even number of Hermitian operators acting on a suitable separable Hilbert space. Inspired by the classical counterpart, one then needs half of these coordinate operators  to be position operators and the remaining ones to be the respective conjugate momenta operators. In the study of quantum phase space (see, \cite{szabo}, for example), the position and momenta operators are taken to be unbounded Hermitian operators representing the noncentral generators of the Weyl-Heisenberg group.  For example, in case of 4-dimensional classical phase space, if one denotes by $x$ and $y$, the position coordinates and by $p_x$ and $p_y$, the respective momenta coordinates, then the corresponding quantum phase space coordinates comprise of the Hermitian operators $\hat{x}$, $\hat{y}$, $\hat{p}_{x}$, $\hat{p}_{y}$ defined on $L^{2}(\mathbb{R}^{2},dx\;dy)$ satisfying the following commutation relations
\begin{align}
&[\hat{x},\hat{p}_{x}]=[\hat{y},\hat{p}_{y}]=i\hbar\mathbb{I},\nonumber\\
&[\hat{x},\hat{y}]=[\hat{p}_{x},\hat{p}_{y}]=0, \label{Heisenberg-algebra}
\end{align}
where $\mathbb{I}$ is the identity operator on $L^{2}(\mathbb{R}^{2},dx\;dy)$. The above commutation relations indeed correspond to the 5-dimensional Weyl-Heisenberg group, abbreviated as $\wh$, in the sequel.

There had been attempts in the past (\cite{gangopadhyay,dulat,ncanistropicoscltr,dulat2009quantum,Dayietal,Girietal,ncqheharmsetal}) where the authors modified the canonical commutation relations (CCR) \eqref{Heisenberg-algebra} by incorporating spatial noncommutativity governed by $[\hat{X},\hat{Y}]=i\vartheta\mathbb{I}$ with the parameter $\vartheta$ taking values in the set of real numbers. Quantum mechanical systems under the influence of uniform magnetic field were studied thoroughly in the past (see, for example, \cite{Landau}). The authors in  \cite{gangopadhyay,dulat,ncanistropicoscltr,dulat2009quantum,Dayietal,Girietal,ncqheharmsetal} and in some articles cited therein mimicked the minimal coupling prescription of quantum mechanics in the noncommutative setting. To this end, the vector potential they come up with is given by
\begin{equation}\label{gauge-lit}
\mathbf{\hat{A}}(\hat{X},\hat{Y})=(-B(1-r)\hat{Y},rB\hat{X}).
\end{equation}
Here the quantum mechanical position operators (multiplication operators that commute with each other) are simply lifted to noncommutative quantum mechanical position operators $\hat{X}$ and $\hat{Y}$ satisfying the commutation relation $[\hat{X},\hat{Y}]=i\vartheta\mathbb{I}$. According to them, $r=1$ and $r=\frac{1}{2}$ correspond to Landau and symmetric gauges, respectively. Later minimal prescription was used rather naively to write down the momentum operators with the help of vector potentials given by \eqref{gauge-lit}:
\begin{equation}
\hat{P}_{i}=\hat{p}_{i}-e\hat{A}_{i}, i=x,y.
\end{equation}
We show in this paper explicitly that such naive minimal prescription yields gauge dependence of the underlying energy spectra for the cases of anisotropic harmonic oscillator (section \ref{sec:landau-symmetric}) and quantum Hall effect (section \ref{sec:Hall Hamiltonian}) in this noncommutative setup. 

Contrary to the above mentioned naive minimal prescription, we relied entirely on families of self-adjoint irreducible representations of the universal enveloping algebra $\mathcal{U}(\G)$ of the Lie algebra $\G$ whose corresponding Lie group $\g$ has been established in an earlier paper \cite{wigfunc} as the kinematical symmetry group of 2-dimensional noncommutative quantum mechanics. A 2-parameter $(r,s)$ family of self-adjoint irreducible representations of $\mathcal{U}(\G)$ is provided in \eqref{eq:algebra-rep-gauge-before-manipulation} of section \ref{sec:group-theory}. It is important to note that despite their appearances in \eqref{eq:algebra-rep-gauge-before-manipulation}, the 2 parameters $r,s$ do not appear in the commutation relations of the algebra, i.e. 

\begin{eqnarray}\label{commut-rel}
&&[\hat{X}^{s}, \hat{\Pi}_{x}^{r,s}]=[\hat{Y}^{s}, \hat{\Pi}_{y}^{r,s}]=i\hbar\mathbb{I},\nonumber\\
&&[\hat{X}^{s},\hat{Y}^{s}]=i\vartheta\mathbb{I},\\
&&[\hat{\Pi}_{x}^{r,s},\hat{\Pi}_{y}^{r,s}]=i\hbar B\mathbb{I}.\nonumber 
\end{eqnarray}

It is, in this sense, we call them {\em gauge parameters} and the 2-parameter family of representations \eqref{eq:algebra-rep-gauge-before-manipulation} as gauge equivalent representations. By tuning the gauge parameters $s\in\mathbb{R}, r\in\mathbb{R}\setminus\{\frac{\hbar}{B\vartheta}\}$, one runs through the same equivalence class of self-adjoint irreducible representations $\mathcal{U}(\G)$ due to the fixed triple $(\frac{1}{\hbar},-\frac{\vartheta}{\hbar^2},-\frac{B}{\hbar})$. Two representatives of this gauge equivalence classes  due to $r=1,s=0$ and $r=\dfrac{\hbar}{\hbar+\sqrt{\hbar(\hbar-\vartheta B)}}, s=\frac{1}{2}$ correspond to the familiar {\em Landau gauge} and {\em symmetric gauge}, respectively.

Using gauge equivalent representatives of the same equivalence class of irreducible self-adjoint representations of $\mathcal{U}(\G)$ given by \eqref{eq:algebra-rep-gauge-before-manipulation}, one writes down the Hamiltonian (see \eqref{eq: Hamiltonian-charged-particle}) of an electron moving on a 2D noncommutative plane immersed in a constant magnetic field under the influence of a polynomial potential. We give a simple general proof towards the end of section \ref{sec:group-theory} on why the spectra of such a Hamiltonian shouldn't depend on the gauge parameters $r,s$. We have later shown in section \ref{group-th cal} how our group theoretic construction indeed yields the same spectra of the underlying Hamiltonian for noncommutative anisotropic harmonic oscillator in both Landau and symmetric gauges. Subsequently, we carry out similar calculations for noncommutative Hall Hamiltonian to obtain its gauge invariant spectra using our group theoretic method in section \ref{gauge independence}. 

The paper is organized as follows. In section \ref{sec:group-theory}, we conduct a short group theoretical discussion of the 7 dimensional Lie group $\g$ along the line of \cite{ncqmjpa}. We also list a few of its important unitary irreducible representations (see \cite{ncqmjpa} for a detailed account) here that will concern us in the following sections. All the representations in section \ref{sec:group-theory} are expressed in terms of the physically relevant parameters $\hbar$, $B$ and $\vartheta$. Later in the section, we provide the 2-parameter $(r,s)$ family of irreducible self-adjoint representations of $\mathcal{U}(\G)$ along the line of \cite{nctori}. We then write down the Hamiltonian of an electron in a noncommutative 2-dimensional plane for a background magnetic field under the influence of a polynomial potential and prove that its spectra do not depend on the parameters $r$ and $s$. We work out the gauge invariant energy spectra for the examples of anisotropic harmonic potential and the Hall potential in section \ref{sec:landau-symmetric} and section \ref{sec:Hall Hamiltonian}, respectively. We also show in the respective sections that the gauge invariance of the underlying energy spectra gets compromised when minimal coupling prescription is applied naively. Section \ref{sec:conclusion} is dedicated to concluding remarks and some possible future research directions.

\section{Group theoretical structure associated with $\g$}\label{sec:group-theory}
For the case of a 2-dimensional quantum mechanical system, the 5-dimensional Weyl-Heisenberg group $\wh$ can be regarded as its kinematical symmetry group. The phase space for an unconstrained 2-dimensional system is $\mathbb{R}^{4}$. The Lie group $\wh$ is just a nontrivial central extension of the underlying abelian group of translations in $\mathbb{R}^4$. Therefore, a generic element of the 5-dimensional Lie group $\wh$ is represented by $(\theta, x, y, p_x, p_y)$ where $\theta$ is the central extension which incorporates the non-commutativity of space and momentum of standard quantum mechanics. Accordingly, the Weyl-Heisenberg Lie algebra, denoted by $\WH$, admits a realization of self-adjoint differential operators on the smooth vectors of $L^2(\mathbb{R}^{2})$; the commutation relations for which are already given in \ref{Heisenberg-algebra}.

This set of commutation relations is known as the canonical commutation relation(CCR). Here, $\hat{x}$, $\hat{y}$, $\hat{p}_{x}$, and $\hat{p}_{y}$ are the self-adjoint representations of the Lie algebra noncentral basis elements on the smooth vectors of $L^{2}(\mathbb{R}^{2})$ with respect to the standard Lebesgue measure. The central basis element of the Lie algebra is mapped to scalar multiple of the identity operator $\mathbb{I}$ on $L^{2}(\mathbb{R}^{2})$.

Since a central extension of the abelian group of translations in $\mathbb{R}^4$ underlies the kinematical symmetry group of standard quantum mechanics by incorporating the space-momentum non-commutativity, it is  worth looking for a triply extended group of translations in $\mathbb{R}^4$ which can appropriately be coined as the kinematical symmetry group of 2-dimensional noncommutative quantum mechanics (NCQM). This was accomplished in \cite{ncqmjpa}. This group, denoted by $\g$, has 3 central elements in contrast to the single central element of $\wh$ so that it can incorporate not only the space-momentum noncommutativity of standard quantum mechanics, but also the space-space and momemtum-momemtum non-commutativity of NCQM (see, for example, the excellent review work \cite{ncqmreview} on NCQM). A generic element of $\g$ will be denoted by $(\theta, \phi, \psi, x, y, p_x, p_y)$ where $x$, $y$ denote the position coordinates and $p_x$, $p_y$ denote the respective momenta. The group composition law for $\g$ is as follows \cite{ncqmjpa}
\begin{align}
&(\theta, \phi, \psi, x, y, p_{x},p_{y})(\theta^\prime, \phi^\prime, \psi^\prime, x^\prime, y^{\prime}, p^{\prime}_{x},p^{\prime}_{y})\nonumber\\
&=\bigg(\theta+\theta^\prime+\dfrac{\alpha}{2}[xp^{\prime}_{x}+yp^{\prime}_{y}-p_{x}x^{\prime}-p_{y}y^{\prime}], \phi+\phi^\prime+\dfrac{\beta}{2}[p_{x}p^{\prime}_{y}-p_{y}p^{\prime}_{x}], \Psi+\Psi^\prime+\dfrac{\gamma}{2}[xy^{\prime}-yx^{\prime}]\nonumber\\ 
&\qquad,x+x^{\prime},y+y^{\prime},p_{x}+p^{\prime}_{x},p_{y}+p^{\prime}_{y}\bigg)\label{eq:group-mult}
\end{align}
where $\alpha, \beta$ and $\gamma$ are certain strictly positive dimensionful constants associated with the central
extensions corresponding to $\theta, \phi$ and $\psi$ respectively. In the case of $\g$, the noncentral generators can be suitably realized as self-adjoint differential operators, viz. $\hat{X}, \hat{Y}, \hat{\Pi}_{x}, \hat{\Pi}_{y}$,  on the space of smooth vectors of $L^2(\mathbb{R}^2)$ obeying the following set of commutation relations:
\begin{eqnarray}\label{commut-rel-without-r-s}
&&[\hat{X}, \hat{\Pi}_{x}]=[\hat{Y}, \hat{\Pi}_{y}]=i\hbar\mathbb{I},\nonumber\\
&&[\hat{X},\hat{Y}]=i\vartheta\mathbb{I},\\
&&[\hat{\Pi}_{x},\hat{\Pi}_{y}]=i\hbar B\mathbb{I},\nonumber 
\end{eqnarray}

with the identification $\hbar=\dfrac{1}{\rho\alpha}$,   $\vartheta=-\dfrac{\sigma\beta}{(\rho\alpha)^2}$ and $B=-\dfrac{\tau\gamma}{\rho\alpha}$ where the ordered triple $(\rho,\sigma,\tau)$ designates an element of the unitary dual $\gd$, i.e. the equivalence classes of the unitary irreducible representations (UIRs) of $\g$. The central generators of $\g$ are all mapped to scalar multiples of the identity operator $\mathbb{I}$ on $L^2(\mathbb{R}^2)$ under the representation \eqref{commut-rel-without-r-s}. Henceforth, we will set the numerical values of the dimensionful constants $\alpha$, $\beta$ and $\gamma$ to 1 and by $\hbar$, $\vartheta$ and $B$, we will only denote their numerical values ignoring their respective dimensions.  

The UIRs of $\g$ and hence the irreducible self-adjoint representations of $\G$ were classified based on the ordered triple $\left(\dfrac{1}{\hbar},-\dfrac{\vartheta}{\hbar^{2}},-\dfrac{B}{\hbar}\right)$. We only list 3 families relevant to our study. The rest can be found in \cite{ncqmjpa}.

\subsection{Case I: $\dfrac{1}{\hbar}\neq 0$, $\dfrac{\vartheta}{\hbar^{2}}\neq 0$, $\dfrac{B}{\hbar}\neq 0$ with $\hbar-\vartheta B\neq 0$}\label{subsec: nondegenerate-rep}
This family of irreducible representations of the universal enveloping algebra $\mathcal{U}(\G)$ realized as self-adjoint differential operators on the smooth vectors of $L^{2}(\mathbb{R}^{2},dx dy)$ is given by
\begin{equation}\label{eq:case1}
\begin{aligned}
&\hat{X}=\hat{x}-\frac{\vartheta}{\hbar}\hat{p}_{y},&\quad& \hat{Y}=\hat{y},\\
&\hat{\Pi}_{x}=\hat{p}_{x}, &\quad&\hat{\Pi}_{y}=-B\hat{x}+\hat{p}_{y},
\end{aligned}
\end{equation}
where the quantum mechanical position and momentum operators are denoted by $\hat{x},\hat{y}$ and $\hat{p}_{x},\hat{p}_{y}$, respectively. They act on smooth vectors $f\in L^{2}(\mathbb{R}^{2},dx\;dy)$ in the following canonical way:
\begin{equation}
\begin{aligned}
(\hat{x}f)(x,y)=xf(x,y),&\quad&(\hat{y}f)(x,y)=yf(x,y),\\
\quad (\hat{p}_{x}f)(x,y)=-i\hbar\dfrac{\partial f}{\partial x}(x,y),&\quad& (\hat{p}_{y}f)(x,y)=-i\hbar\dfrac{\partial f}{\partial y}(x,y).\label{eq: action-QM-operators}
\end{aligned}
\end{equation} 
\begin{figure}[h!]
\centerline{\includegraphics[width=15cm]{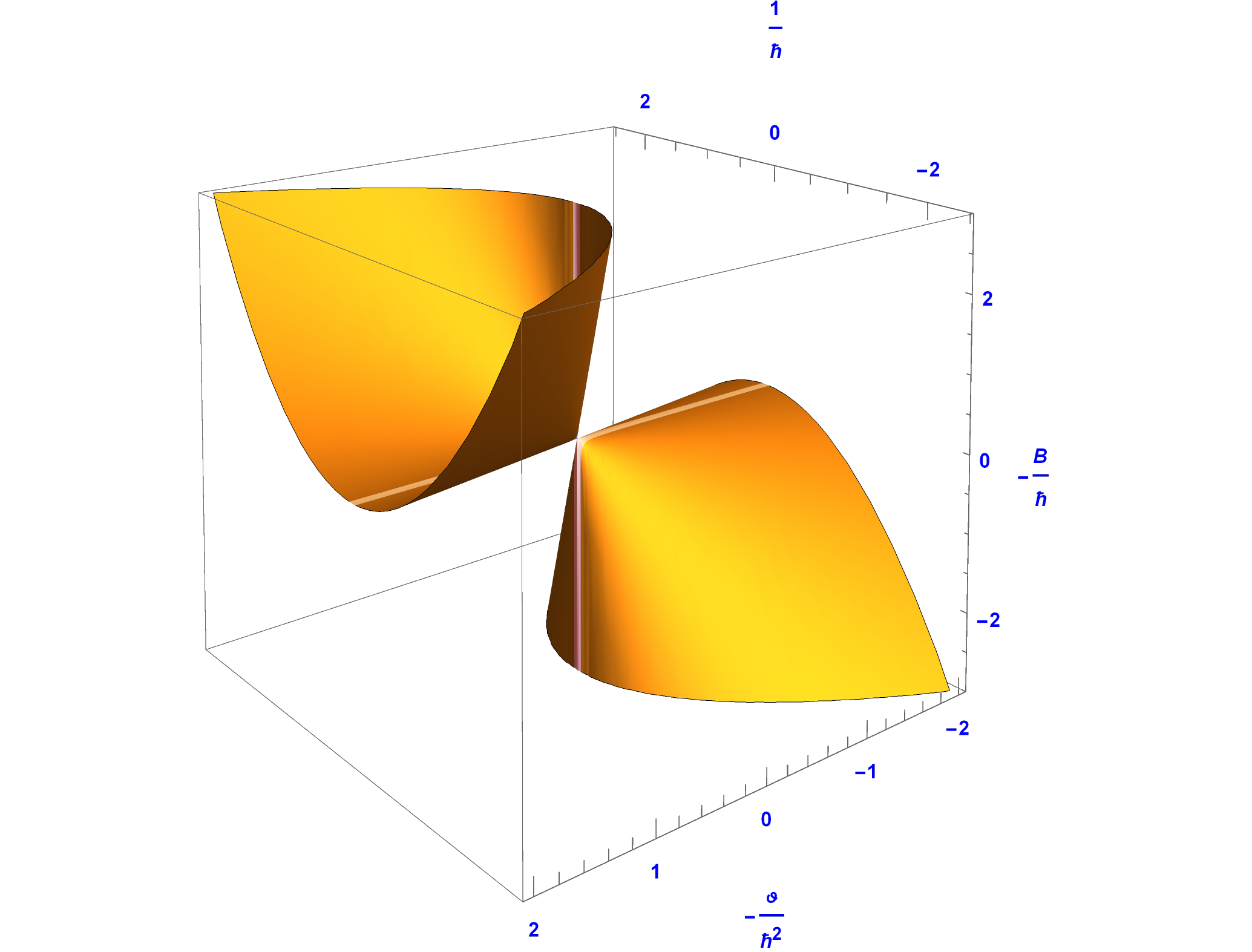}}\vspace{0cm}
\caption{The surface $\mathbb{S}_{\frac{1}{\hbar},\zeta}$ in $\mathbb{R}_0^3$ that is associated with a family of two-dimensional coadjoint orbits in the dual Lie algebra $\G^*$. Inside $\mathbb{R}_0^3$, is embedded an elliptic cone-shaped surface (with two perpendicular lines deleted) given by the equation $\frac{1}{\hbar^{2}}-\frac{\vartheta B}{\hbar^{3}}= 0$. Any point on such a surface is completely determined by a family of straight lines given by $\frac{1}{\hbar}=-\frac{\vartheta\zeta}{\hbar^{2}}=-\frac{B}{\zeta\hbar}$ with $\zeta \in (-\infty,0) \cup (0, \infty)$.}
\label{lanfig}
\end{figure}

\subsection{Case II: $\dfrac{1}{\hbar}\neq 0$, $\dfrac{\vartheta}{\hbar^{2}}\neq 0$, $\dfrac{B}{\hbar}\neq 0$ with $\hbar-\vartheta B=0$}
This family of irreducible representations of $\mathcal{U}(\G)$ realized as self-adjoint differential operators on $L^{2}(\mathbb{R})$ is given as follows
\begin{equation}\label{eq:degenerate-rep}
\begin{aligned}
&\hat{X}=-\hat{x},&\quad& \hat{Y}=-\dfrac{\vartheta}{\hbar}\hat{p},\\
&\hat{\Pi}_{x}=\hbar\kappa\mathbb{I}-\hat{p}, &\quad&\hat{\Pi}_{y}=\hbar\delta\mathbb{I}+\frac{\hbar}{\vartheta}\hat{x},
\end{aligned}
\end{equation}
where the self-adjoint unbounded operators $\hat{x}$ and $\hat{p}$ on $L^{2}(\mathbb{R},dx)$ are given in terms of their actions on smooth vectors as
\begin{equation}
(\hat{x}f)(x)=xf(x),\quad (\hat{p}f)(x)=-i\hbar\frac{\partial f}{\partial x}(x),
\end{equation}
so that, indeed, $[\hat{x},\hat{p}]=i\hbar\mathbb{I}$ with $\mathbb{I}$ being the identity operator on $L^{2}(\mathbb{R},dx)$. Here, the 2 independent parameters $\kappa$ and $\delta$ label the foliations of 2-dimensional Euclidean planes inside the 4-dimensional ones. The details can be found in \cite{ncqmjpa}. For each quadruple $(\frac{1}{\hbar},-\frac{B}{\hbar},\kappa,\delta)$, one finds an irreducible self-adjoint representation of the universal enveloping algebra $\mathcal{U}(\G)$ in $L^{2}(\mathbb{R},dx)$ as give by \eqref{eq:degenerate-rep}.

It is interesting to note that there exists a family of UIRs of the Lie algebra $\G$ which obey the canonical commutation relation(CCR) of quantum mechanics. This case is considered below:

\subsection{Case III: $\dfrac{1}{\hbar}\neq 0$, $\vartheta=0$, $B=0$}
Here, the irreducible representation of the universal enveloping algebra $\mathcal{U}(\G)$ is realized as self-adjoint differential operators on the smooth vectors of $L^{2}(\mathbb{R}^{2},dxdy)$ in the following way
\begin{equation}\label{eq: QM-rep}
\begin{aligned}
&\hat{X}=\hat{x},&\quad& \hat{Y}=\hat{y},\\
&\hat{\Pi}_{x}=\hat{p}_{x}, &\quad&\hat{\Pi}_{y}=\hat{p}_{y},
\end{aligned}
\end{equation}
where the actions of the unbounded self-adjoint operators $\hat{x}$, $\hat{y}$, $\hat{p}_{x}$ and $\hat{p}_{y}$ on $f\in L^{2}(\mathbb{R}^{2},dxdy)$ are given by \eqref{eq: action-QM-operators}. It is, therefore, sufficient to resort to the group $\g$ to obtain the CCR. However, it should be emphasized that $\wh$ is not a subgroup of $\g$.

Another interesting point, which is more relevant to this paper, is that two certain gauge equivalent representations of NCQM, viz., the Landau and the symmetric gauge representations, arise from two equivalent irreducible self-adjoint representations of $\mathcal{U}(\G)$ determined by a fixed value of $\dfrac{1}{\hbar}\neq 0$, $\vartheta\neq 0$, and $B\neq 0$ satisfying $\hbar-B\vartheta\neq 0$. The underlying equivalence class is precisely the one given by Case \ref{subsec: nondegenerate-rep} in the list above due to fixed ordered triple $\left(\dfrac{1}{\hbar},-\dfrac{\vartheta}{\hbar^{2}},-\dfrac{B}{\hbar}\right)$. In other words, different values of the ordered pair $(r,s)$ will always yield equivalent irreducible representations belonging to the same equivalence class. In the rest of this section, we discuss this family of equivalent irreducible self-adjoint representations of the universal enveloping algebra $\mathcal{U}(\G)$; a family to which Landau and symmetric gauge representations belong.

This 2-parameter family of equivalent self-adjoint irreducible representation of the universal enveloping algebra $\mathcal{U}(\G)$ on the smooth vectors of $L^{2}(\mathbb{R}^{2},dxdy)$ is given as follows:
\begin{equation}\label{eq:algebra-rep-gauge-before-manipulation}
\begin{aligned}
 \hat{X}^s&=\hat{x}-s\dfrac{\vartheta}{\hbar}\hat{p}_{y},\\
 \hat{Y}^s&=\hat{y}+(1-s)\dfrac{\vartheta}{\hbar}\hat{p}_{x},\\
 \hat{\Pi}_{x}^{r,s}&=\dfrac{(1-r)\hbar B}{\hbar-r\vartheta B}\hat{y}+\dfrac{[(r+s-rs)\vartheta B-\hbar]}{r\vartheta B-\hbar}\hat{p}_{x},\\
 \hat{\Pi}_{y}^{r,s}&=-rB\hat{x}+\left[1+r(s-1)\dfrac{\vartheta B}{\hbar}\right]\hat{p}_{y},
\end{aligned}
\end{equation}
which, following a brief algebraic manipulation, yields
\begin{equation}\label{eq:algebra-rep-gauge}
\begin{aligned}
 \hat{X}^s&=\hat{x}-s\dfrac{\vartheta}{\hbar}\hat{p}_{y},\\
 \hat{Y}^s&=\hat{y}+(1-s)\dfrac{\vartheta}{\hbar}\hat{p}_{x},\\
 \hat{\Pi}_{x}^{r,s}&=\dfrac{(1-r)\hbar B}{\hbar-r\vartheta B}\left(\hat{y}-\frac{s\vartheta}{\hbar}\hat{p}_{x}\right)+\hat{p}_{x},\\
 \hat{\Pi}_{y}^{r,s}&=-rB\left[\hat{x}+\frac{(1-s)\vartheta}{\hbar}\hat{p}_{y}\right]+\hat{p}_{y}.
\end{aligned}
\end{equation}

Note that $r=1$ and $s=1$ in \eqref{eq:algebra-rep-gauge-before-manipulation} yield a representative of the equivalence class of irreducible self-adjoint representations of the universal enveloping algebra $\mathcal{U}(\G)$ given by case I in \eqref{eq:case1}. At this stage, upon looking at the expressions \eqref{eq:algebra-rep-gauge} of the 2-parameter family of the kinematical momenta $\hat{\Pi}^{r,s}_{x}$ and $\hat{\Pi}^{r,s}_{y}$, one immediately deduces that minimal prescription fails in this noncommutative setting. The terms in parenthesis appearing in the kinematical momenta expressions in \eqref{eq:algebra-rep-gauge} are given by
\begin{equation}\label{eq:kinematical-momenta}
\begin{aligned}
\hat{Y}^{s+1}&=\hat{y}-\frac{s\vartheta}{\hbar}\hat{p}_{x},\\
\hat{X}^{s-1}&=\hat{x}+\frac{(1-s)\vartheta}{\hbar}\hat{p}_{y},
\end{aligned}
\end{equation}
which actually belong to 2 different representation classes $(r, s+1)$ and $(r, s-1)$ (see \eqref{eq:algebra-rep-gauge-before-manipulation}). Hence, there exist no canonical vector potential for the case of noncommutative space in contrast to the standard quantum mechanical setting in the presence of a magnetic field. A definition of vector potentials in such a noncommutative setting as operator valued 1-forms is suggested in \cite{nctori}. This definition is, however, not based on minimal prescription. Agreement of such definitions with Conne's noncommutative geometric prescription is yet to be validated. We, therefore, utilise the explicit expressions of the 2-parameter family of kinematical momenta \eqref{eq:algebra-rep-gauge-before-manipulation} to delineate various gauges involved here. The gauge parameter $s$ can take any value from the real line $\mathbb{R}$ while the allowed values of $r$ are given by $r \in \mathbb{R} \smallsetminus\bigg\lbrace\dfrac{\hbar}{B\vartheta}\bigg\rbrace $.

The Landau and symmetric gauges belong to this gauge equivalent family of irreducible self-adjoint representations of $\G$ for the following parametric values:
\[
r=1, s=0 \quad \textrm{ for Landau gauge},
\]
\[
r=\dfrac{\hbar}{\hbar+\sqrt{\hbar(\hbar-\vartheta B)}}:=r_{\hbox{\tiny{sym}}}, s=\dfrac{1}{2} \quad \textrm{ for symmetric gauge.}
\]

The goal of this paper is to compute the energy spectra of an electron moving in a 2-dimensional noncommutative plane immersed in a constant vertical magnetic field for both the cases of anisotropic harmonic potential and the Hall potential. In fact, we show that the gauge invariance of the energy spectra is maintained for a general class of potentials, namely polynomial potentials. The 2 cases to be studied are examples of polynomial potentials.  For a given ordered triple $\left(\dfrac{1}{\hbar},-\dfrac{\vartheta}{\hbar^{2}},-\dfrac{B}{\hbar}\right)$ and $r\in\mathbb{R}\smallsetminus\bigg\lbrace\dfrac{\hbar}{B\vartheta}\bigg\rbrace, s\in\mathbb{R}$, the Hamiltonian for such a charged particle (the charge $e$ is conveniently set to unity) under a general polynomial potential $V(\hat{X}^{s},\hat{Y}^{s})$ is given by
\begin{equation}\label{eq: Hamiltonian-charged-particle}
\hat{H}^{r,s}=\frac{1}{2m}[(\hat{\Pi}^{r,s}_{x})^{2}+(\hat{\Pi}^{r,s}_{y})^{2}]+V(\hat{X}^{s},\hat{Y}^{s}),
\end{equation}
with $m$ being the mass of the charged particle. The 2-parameter $(r,s)$ family of unitary irreducible representations of $\g$ associated with the fixed ordered triple $(\hbar,\vartheta,B)$ all belong to the same equivalence class of the unitary dual $\gd$. Two such representations labeled by $(r,s)$ and $(r^{\prime},s^{\prime})$ are intertwined by a unitary operator $U$ on the given Hilbert space $L^{2}(\mathbb{R}^{2},dx\;dy)$. Consequently, the group generators also transform, using the same unitary operator $U$, as follows
\begin{equation}\label{eq:generators-unitary_transformation}
\begin{aligned}
&\hat{\Pi}_{x}^{r^{\prime},s^{\prime}}=U\hat{\Pi}^{r,s}_{x}U^{-1},\\
&\hat{\Pi}_{y}^{r^{\prime},s^{\prime}}=U\hat{\Pi}^{r,s}_{y}U^{-1},\\
&\hat{X}^{s^{\prime}}=U\hat{X}^{s}U^{-1},\\
&\hat{Y}^{s^{\prime}}=U\hat{Y}^{s}U^{-1},
\end{aligned}
\end{equation}
leading to the following unitary transformation of the underlying Hamiltonian: 
\begin{equation}\label{eq:Hamiltonian-unitary-transformation}
\hat{H}^{r^{\prime},s^{\prime}}=U\hat{H}^{r,s}U^{-1}.
\end{equation}
One writes down the Schr\"{o}dinger equation for the Hamiltonian $H^{r,s}$ having eigenfunction $\psi^{r,s}\in L^{2}(\mathbb{R}^{2},dx\;dy)$ with eigenvalue $E$, as
\begin{equation}\label{eq:schroedinger-equation}
\hat{H}^{r,s}\psi^{r,s}=E\psi^{r,s},
\end{equation}
which, after acted upon by $U$ from the left on both sides, yields
\begin{equation*}
U\hat{H}^{r,s}\psi^{r,s}=EU\psi^{r,s}.
\end{equation*}
The above equation can be rearranged to yield
\begin{equation*}
U\hat{H}^{r,s}U^{-1}(U\psi^{r,s})=E(U\psi^{r,s}).
\end{equation*}
In other words, one obtains,
\begin{equation*}\label{eq:transformed-Schroedinger-equation}
\hat{H}^{r^{\prime},s^{\prime}}\tilde{\psi}^{r^{\prime},s^{\prime}}=E\tilde{\psi}^{r^{\prime},s^{\prime}},
\end{equation*}
where $\tilde{\psi}^{r^{\prime},s^{\prime}}=U\psi^{r,s}$ is the eigenfunction of the unitarily transformed Hamiltonian $\hat{H}^{r^{\prime},s^{\prime}}$ (see \eqref{eq:Hamiltonian-unitary-transformation}) with the same eigenvalue $E$.

Therefore, the spectra of the underlying Hamiltonian operator is independent of the gauge parameters $r$ and $s$. In what follows we discuss 2 physically important examples of polynomial potentials, namely anisotropic harmonic potential and Hall potential in the light of the above discussion on gauge invariant energy spectra.

\section{Noncommutative two dimensional anisotropic harmonic oscillator in a constant magnetic field}\label{sec:landau-symmetric}
In this section, we consider the noncommutative Landau problem subjected to an anisotropic harmonic potential defined on the plane to demonstrate the gauge independence of energy eigenvalues.

For the case of noncommutative Landau problem, the quantized phase space coordinates, i.e. self-self-adjointadjoint unbounded operators on $L^{2}(\mathbb{R}^{2},dx\;dy)$ that represent the classical position and momentum coordinates, obey the following commutation relations 
\begin{equation}\label{commutation-rescaled}
[\hat{X},\hat{Y}]=i\vartheta\mathbb{I},\,\,\,[\hat{\Pi}_x,\hat{\Pi}_y]=i\hbar B\mathbb{I}\,\,\,\text{and}\,\,\, [\hat{X},\hat{\Pi}_x]=[\hat{Y},\hat{\Pi}_y]=i\hbar\mathbb{I},
\end{equation} 
where $\mathbb{I}$ is the identity operator on $L^{2}(\mathbb{R}^{2}, dx\;dy)$. Here, note that the magnetic field $B$ can be rescaled $B\rightarrow\frac{eB}{c}$ to connect our notation with the usual literature on Landau problem. Moreover, we define the Cyclotron frequency to be
\[
\omega_c=\frac{B}{m}.
\]
\subsection{Energy eigenvalue calculation}\label{enegvalue} 
The Hamiltonian describing the noncommutative Landau problem under the influence of anisotropic harmonic potential can be obtained by substituting $V(\hat{X}^{s},\hat{Y}^{s})=\frac{1}{2}m\big[\omega_1^2(\hat{X}^{s})^2+\omega_2^2(\hat{Y}^{s})^2\big]$ in \eqref{eq: Hamiltonian-charged-particle} which is as follows
\begin{equation}\label{eq:NC-Hamiltonian}
H^{NC}=\frac{1}{2m}\big[(\hat{\Pi}_{x}^{r,s})^2+(\hat{\Pi}_y^{r,s})^2\big]+\frac{1}{2}m\big[\omega_1^2(\hat{X}^{s})^2+\omega_2^2(\hat{Y}^{s})^2\big].
\end{equation}

Now to calculate its energy spectrum we need to use quantum mechanical operators ${\hat{x},\hat{y},\hat{p}_{x},\hat{p}_{y}}$ on $L^{2}(\mathbb{R}^{2},dx\;dy)$. Using \eqref{eq:algebra-rep-gauge-before-manipulation} for symmetric gauge $(r=r_{\text{Sym}},\,s=1/2)$ or Landau gauge $(r=1,\, s=0)$, we obtain the following form of the Hamiltonian:
\begin{equation}\label{eq:3.3}
\begin{aligned}
H=\frac{1}{2M_1}\hat{p}_{x}^2+\frac{1}{2M_2}\hat{p}_{y}^2+\frac{1}{2}M_1\Omega_1^2\hat{x}^2+\frac{1}{2}M_2\Omega_2^2\hat{y}^2-l_1\hat{x}\hat{p}_{y}+l_2\hat{y}\hat{p}_{x}.
\end{aligned}
\end{equation}

The effect of using different gauges and initial parameters of the NC Hamiltonian \eqref{eq:NC-Hamiltonian}, $\{m,\omega_{i},\omega_{c},\vartheta\}$ enter into the new parameters, $\left(M_{i},\Omega_{i},l_{i}\right)$ of \eqref{eq:3.3} which we will present explicitly in subsequent sections. As the Hamiltonian \eqref{eq:3.3} is quadratic in the quantum mechanical position and momentum operators, we redefine \eqref{eq:3.3} in terms of creation and annihilation operators to determine the energy spectrum:

\begin{equation}
\begin{aligned}
\hat{x}&=\sqrt{\frac{\hbar}{2M_1\Omega_1}}\left(\hat{a}_{x}+\hat{a}_{x}^\dagger\right),\,\,\, \hat{p}_{x}=-i\sqrt{\frac{\hbar M_1\Omega_1}{2}}\left(\hat{a}_{x}-\hat{a}_{x}^\dagger\right)\\
\hat{y}&=\sqrt{\frac{\hbar}{2M_2\Omega_2}}\left(\hat{a}_{y}+\hat{a}_{y}^\dagger\right),\,\,\,\hat{p}_{y}=-i\sqrt{\frac{\hbar M_2\Omega_2}{2}}\left(\hat{a}_{y}-\hat{a}_{y}^\dagger\right)
\end{aligned}
\end{equation}
with

\begin{equation}
\begin{aligned}
[\hat{a}_{x},\hat{a}_{x}^\dagger]&=[\hat{a}_{y},\hat{a}_{y}^\dagger]=\mathbb{I},\\
[\hat{a}_{x},\hat{a}_{y}]&=[\hat{a}_{x}^\dagger,\hat{a}_{y}^\dagger]=[\hat{a}_{x},\hat{a}_{y}^\dagger]=[\hat{a}_{y},\hat{a}_{x}^\dagger]=0.
\end{aligned}
\end{equation}

Therefore, the Hamiltonian \eqref{eq:3.3} now becomes,
\begin{equation}\label{eq:17}
H=\frac{\hbar\Omega_1}{2}\big(\hat{a}_{x}\hat{a}_{x}^\dagger+\hat{a}_{x}^\dagger \hat{a}_{x}\big)+\frac{\hbar\Omega_2}{2}\big(\hat{a}_{y}\hat{a}_{y}^\dagger+\hat{a}_{y}^\dagger \hat{a}_{y}\big)
+\frac{i\hbar}{2}\big[c(\hat{a}_{x}\hat{a}_{y}-\hat{a}_{x}^\dagger \hat{a}_{y}^\dagger)+d(\hat{a}_{x}^\dagger \hat{a}_{y}-\hat{a}_{y}^\dagger \hat{a}_{x})\big]
\end{equation}\\
where,
\begin{equation}
c=l_1\sqrt{\frac{M_2\Omega_2}{M_1\Omega_1}}-l_2\sqrt{\frac{M_1\Omega_1}{M_2\Omega_2}},\,\,\,d=l_1\sqrt{\frac{M_2\Omega_2}{M_1\Omega_1}}+l_2\sqrt{\frac{M_1\Omega_1}{M_2\Omega_2}}.
\end{equation}
\\

If the parameter $l_{1}$ of \eqref{eq:3.3} becomes negative, the parameters $c$ and $d$ of \eqref{eq:17} are interchanged i.e. $c\rightarrow d$ and $d\rightarrow c$ with $l_{1}\rightarrow |l_{1}|$.

By applying the diagonalization method for general bosonic bilinear Hamiltonian \cite{tsallis, maldonado}, we determine from \eqref{eq:17} the following diagonal Hamiltonian,
\begin{equation}\label{eq:dig}
H^{Dig}= \frac{1}{2}\hbar\tilde{\Omega}_1(\hat{B}_{1}\hat{B}_{1}^\dagger +\hat{B}_{1}^\dagger\hat{B}_{1})+ \frac{1}{2}\hbar\tilde{\Omega}_{2}(\hat{B}_{2}\hat{B}_{2}^\dagger +\hat{B}_{2}^\dagger\hat{B}_{2}),
\end{equation}
where the eigenfrequencies $\tilde{\Omega}_{1,2}$ are
\begin{equation}\label{eq:egfreq}
\tilde{\Omega}_{1,2}=\left[\frac{C_{1}}{2}\pm \sqrt{\frac{C_{1}^{2}}{4}-C_{2}}\right]^{1/2}\!\!\!\!\!\!\!\!,
\end{equation}
with
\begin{align}
C_{1}&=\hbar^2(\Omega_{1}^{2}+\Omega_{2}^{2}-c^2/2+d^2/2),\nonumber\\
C_{2}&= \hbar^4\left[\Omega_{1}^{2}\Omega_{2}^{2}- \Omega_{1}\Omega_{2}(c^2+d^2)/2+(c^{2}-d^{2})^{2}/4\right].\nonumber
\end{align}
Consequently, we see that the diagonalized Hamiltonian decomposes into two independent pieces of one dimensional Harmonic oscillator with frequencies $\tilde{\Omega}_{1}$ and $\tilde{\Omega}_{2}$, respectively. Moreover, \eqref{eq:dig} can be written as
\begin{equation}\label{eq:num}
H^{Dig}=\hbar\tilde{\Omega}_{1}\left(\hat{N}_{1}+\frac{1}{2}\mathbb{I} \right)+\hbar\tilde{\Omega}_{2}\left(\hat{N}_{2}+\frac{1}{2}\mathbb{I}\right),
\end{equation}
where the number operators are defined as $\hat{N}_{1}=\hat{B}_{1}^{\dagger}\hat{B}_{1}$ and $\hat{N}_{2}=\hat{B}_{2}^{\dagger}\hat{B}_{2}$. Therefore, the energy eigenvalues read
\begin{equation}\label{eq:energy}
E_{n_{1},n_{2}}=\hbar\tilde{\Omega}_{1}\left(n_{1}+\frac{1}{2}\right)+\hbar\tilde{\Omega}_{2}\left(n_{2}+\frac{1}{2}\right).
\end{equation}

\subsection{Group theoretic Construction}\label{group-th cal}

The transformation between NCQM operators and quantum mechanical operators on the Hilbert space $L^{2}(\mathbb{R}^{2},dx\;dy)$ is achieved by means of representation theory. For the symmetric gauge, the transformation is as follows:\\
\begin{equation}
\begin{aligned}
\hat{X}_{\text{Sym}}&=\hat{x}-\frac{\vartheta}{2\hbar}\hat{p}_y,\,\,\,\hat{Y}_{\text{Sym}}=\hat{y}+\frac{\vartheta}{2\hbar}\hat{p}_x\\
\hat{\Pi}_{x_{\text{Sym}}}&=\frac{\hbar B}{\hbar+\sqrt{\hbar(\hbar-\vartheta B)}}\hat{y}+\frac{\hbar+\sqrt{\hbar(\hbar-\vartheta B)}}{2\hbar}\hat{p}_x,\\
\hat{\Pi}_{y_{\text{Sym}}}&=-\frac{\hbar B}{\hbar+\sqrt{\hbar(\hbar-\vartheta B)}}\hat{x}+\frac{\hbar+\sqrt{\hbar(\hbar-\vartheta B)}}{2\hbar}\hat{p}_y.
\end{aligned}
\end{equation} 
Then, the Hamiltonian $H^{NC}$, in terms of quantum mechanical operators ${\hat{x},\hat{y},\hat{p}_{x},\hat{p}_{y}}$ reads as
\begin{equation}\label{eq:10}
\begin{aligned}
H=\frac{1}{2M_1}\hat{p}_{x}^2+\frac{1}{2M_2}\hat{p}_{y}^2+\frac{1}{2}M_1^2\Omega_1^2\hat{x}^2+\frac{1}{2}M_2\Omega_2^2\hat{y}^2-l_1\hat{x}\hat{p}_{y}+l_2\hat{y}\hat{p}_{x},
\end{aligned}
\end{equation}
where the parameters appearing in eq.(\ref{eq:10}) are given by
\begin{equation}\label{eq:sym}
\begin{aligned}
M_{1_{\text{Sym}}}&=\frac{m}{\frac{1}{2}+\frac{m^2\vartheta^2\omega_2^2}{4\hbar^2}-\frac{m\omega_c\vartheta}{4\hbar}+\frac{\sqrt{\hbar(\hbar-m\omega_c\vartheta)}}{2\hbar}},\,\,\,M_{2_{\text{Sym}}}=\frac{m}{\frac{1}{2}+\frac{m^2\vartheta^2\omega_1^2}{4\hbar^2}-\frac{m\omega_c\vartheta}{4\hbar}+\frac{\sqrt{\hbar(\hbar-m\omega_c\vartheta)}}{2\hbar}},\\
{\Omega}^2_{1_{\text{Sym}}}&=\frac{m}{M_{1_{\text{Sym}}}}\bigg[\omega_1^2+\frac{\omega_c^2\hbar^2}{(\hbar+\sqrt{\hbar(\hbar-m\omega_c\vartheta)})^2} \bigg],\,\,\,{\Omega}^2_{2_{\text{Sym}}}=\frac{m}{M_{2_{\text{Sym}}}}\bigg[\omega_2^2+\frac{\omega_c^2\hbar^2}{(\hbar+\sqrt{\hbar(\hbar-m\omega_c\vartheta)})^2} \bigg],\\
l_{1_{\text{Sym}}}&=\frac{\omega_c}{2}+\frac{m\omega_1^2\vartheta}{2\hbar},\,\,\,\,\,\,
l_{2_{\text{Sym}}}=\frac{\omega_c}{2}+\frac{m\omega_2^2\vartheta}{2\hbar}.
\end{aligned}
\end{equation}

On the other hand, for the Landau gauge, we have the following transformation:
\begin{equation}\label{eq:lan}
\begin{aligned}
\hat{X}_{\text{Lan}}&=\hat{x},\,\,\,\hat{Y}_{\text{Lan}}=\hat{y}+\frac{\vartheta}{\hbar} \hat{p}_{x},\\
\hat{\Pi}_{x_{\text{Lan}}}&=\hat{p}_{x},\,\,\,\hat{\Pi}_{y_{\text{Lan}}}=\frac{\hbar-\vartheta B}{\hbar}\hat{p}_{y}-B\hat{x}.
\end{aligned}
\end{equation}\\
In this case, the parameters involved in the Hamiltonian \eqref{eq:10} are given by
\begin{equation}
\begin{aligned}
M_{1_{\text{Lan}}}&=\frac{m}{1+\frac{m^2\vartheta^2\omega_2^2}{\hbar^2}},\,\,\,M_{2_{\text{Lan}}}=\frac{m}{\big(\frac{\hbar-m\omega_c\vartheta}{\hbar}\big)^2},\\
\Omega^{2}_{1_{\text{Lan}}}&=\frac{m}{M_{1_{\text{Lan}}}}\big(\omega_1^2+\omega_c^2\big),\,\,\,\Omega^{2}_{2_{\text{Lan}}}=\frac{m}{M_{2_{\text{Lan}}}}\omega_2^2,\\
l_{1_{\text{Lan}}}&=\omega_c-\frac{m\omega_c^2\vartheta}{\hbar},\,\,\,l_{2_{\text{Lan}}}=\frac{m\omega_2^2\vartheta}{\hbar}.
\end{aligned}
\end{equation}

\begin{figure}[h!]
\centerline{\includegraphics[width=14cm]{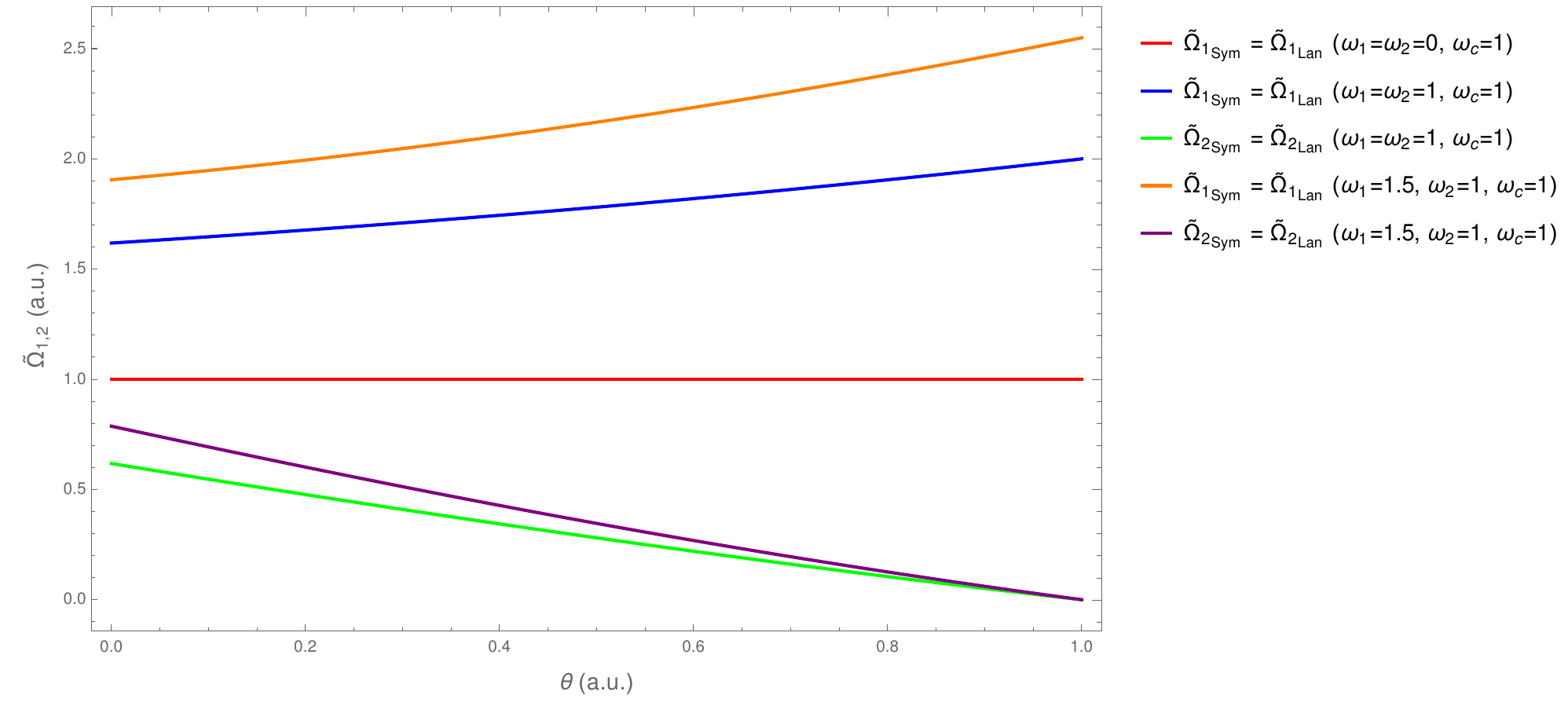}}\vspace{0cm}
\caption{Eigenfrequencies $\tilde{\Omega}_{1,2}$ as a function of $\vartheta$ in arbitrary unit (a.u.) in case of pure Landau ($\omega_{1}=\omega_{2}=0$), isotropic harmonic ($\omega_{1}=\omega_{2}=1$ a.u.) and anisotropic harmonic ($\omega_{1}=1.5, \omega_{2}=1$ a.u.) potentials for both symmetric (Sym) and Landau (Lan) gauges. Here, $\hbar=1$, $m=1$ and $\omega_{c}=1$ in arbitrary unit. The parameter $\vartheta$ is bounded by $\vartheta\leq \frac{\hbar}{m\omega_{c}}$ for the symmetric gauge.}
\label{symfig}
\end{figure}

\begin{figure}[h!]
\centerline{\includegraphics[width=14cm]{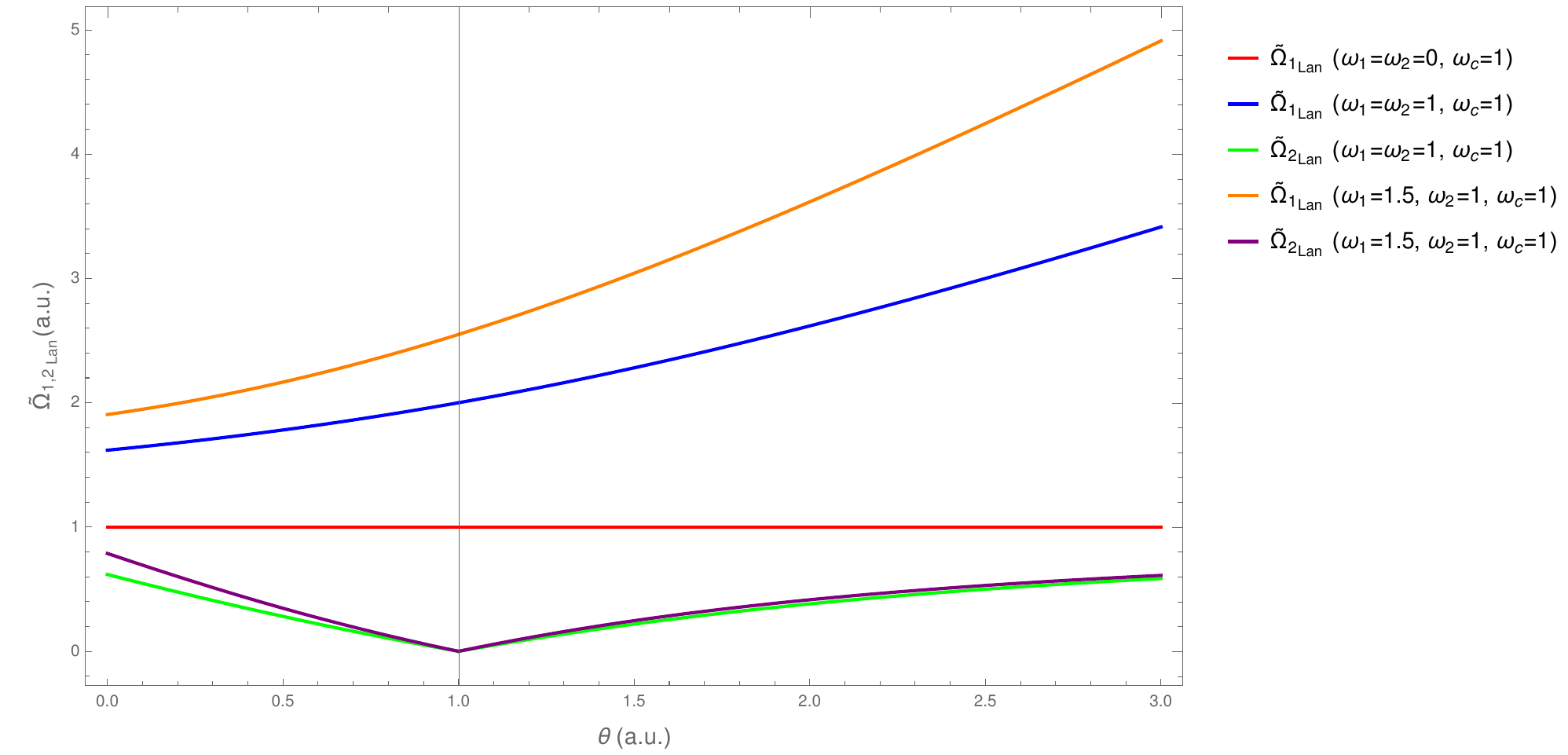}}\vspace{0cm}
\caption{Eigenfrequencies $\tilde{\Omega}_{1,2}$ as a function of $\vartheta$ in arbitrary unit (a.u.) for the Landau gauge in case of pure Landau ($\omega_{1}=\omega_{2}=0$), isotropic harmonic ($\omega_{1}=\omega_{2}=1$ a.u.) and anisotropic harmonic ($\omega_{1}=1.5, \omega_{2}=1$ a.u.) potentials where $\vartheta$ is extended beyond $\frac{\hbar}{m\omega_{c}}$. Here also, $\hbar=1$, $m=1$ and $\omega_{c}=1$ in arbitrary unit.}
\label{lanfig}
\end{figure}

We can see from Figure \ref{symfig} and \ref{lanfig} that the eigenfrequencies $\tilde{\Omega}_{1,2}$ are both gauge invariant quantities. In Fig. \ref{symfig}, $\vartheta$ is constrained to satisfy $\vartheta < \frac{\hbar}{m\omega_{c}}$ to ensure that the parameters given in \eqref{eq:sym} for the symmetric gauge stays real and the Hamiltonian given by \eqref{eq:10} remains self-adjoint. On the other hand, there is no constraint on $\vartheta$ for the Landau gauge to keep the parameters given in \eqref{eq:lan} real. Hence, in Fig. \ref{lanfig}, $\vartheta$ is taken to lie in the domain $0 \leq \vartheta$.  Besides, in Fig. \ref{symfig} and \ref{lanfig}, we can see that the eigenfrequency $\tilde{\Omega}_{2}$ goes to zero for $\vartheta = \frac{\hbar}{m\omega_{c}}$ which signals that the Hilbert space $L^{2}(\mathbb{R}^{2}, dx\;dy)$ degenerates into $L^{2}(\mathbb{R}, dx\;dy)$ in such situation. In addition, we have seen that for the pure Landau problem ( $\omega_{1}=\omega_{2}=0$) the eigenfrequencies, $\tilde{\Omega}_{2}$ becomes zero and $\tilde{\Omega}_{1}=\omega_{c}$, and it does not depend on the $\vartheta$ even if we are considering the noncommutative version of the problem. The reason behind the $\vartheta$ independence of the eigenfrequencies is the following. For the pure Landau problem, the noncommutative Hamiltonian is
\begin{equation}\label{eq:NC-Hamiltonian-Landau}
H^{NC}=\frac{1}{2m}\big[(\hat{\Pi}_{x}^{r,s})^2+(\hat{\Pi}_y^{r,s})^2\big].
\end{equation}
One can rescale the kinematical momenta, for example, $\hat{\Pi}_{y}\rightarrow \hat{Q}=-\hat{\Pi}_{y}/B$ and $\hat{\Pi}_{x}\rightarrow \hat{P}=\hat{\Pi}_{x}$ so that  they form the canonical pair, i.e, their commutation relation, $[\hat{\Pi}_{x},\hat{\Pi}_{y}]=i\hbar B\mathbb{I}$ given in \eqref{commutation-rescaled} now becomes the canonical commutation relation, $[\hat{Q},\hat{P}]=i\hbar\mathbb{I}$ upon rescaling. Therefore, in the absence of any function depending on the position operators, $\hat{X}$ and $\hat{Y}$ in the Hamiltonian, the dynamics will be dictated by the canonical observables, $\hat{Q}$ and $\hat{P}$. As there is no $\vartheta$ dependence in the commutation relation of the kinematical momenta within the group-theoretic construction, the quantum system described by the Hamiltonian given in \eqref{eq:NC-Hamiltonian-Landau} will resemble the one-dimensional harmonic oscillator problem of the standard Quantum Mechanics.

\subsection{Naive Minimal Prescription}\label{naive-minimal}

\begin{figure}[h!]
\centerline{\includegraphics[width=14cm]{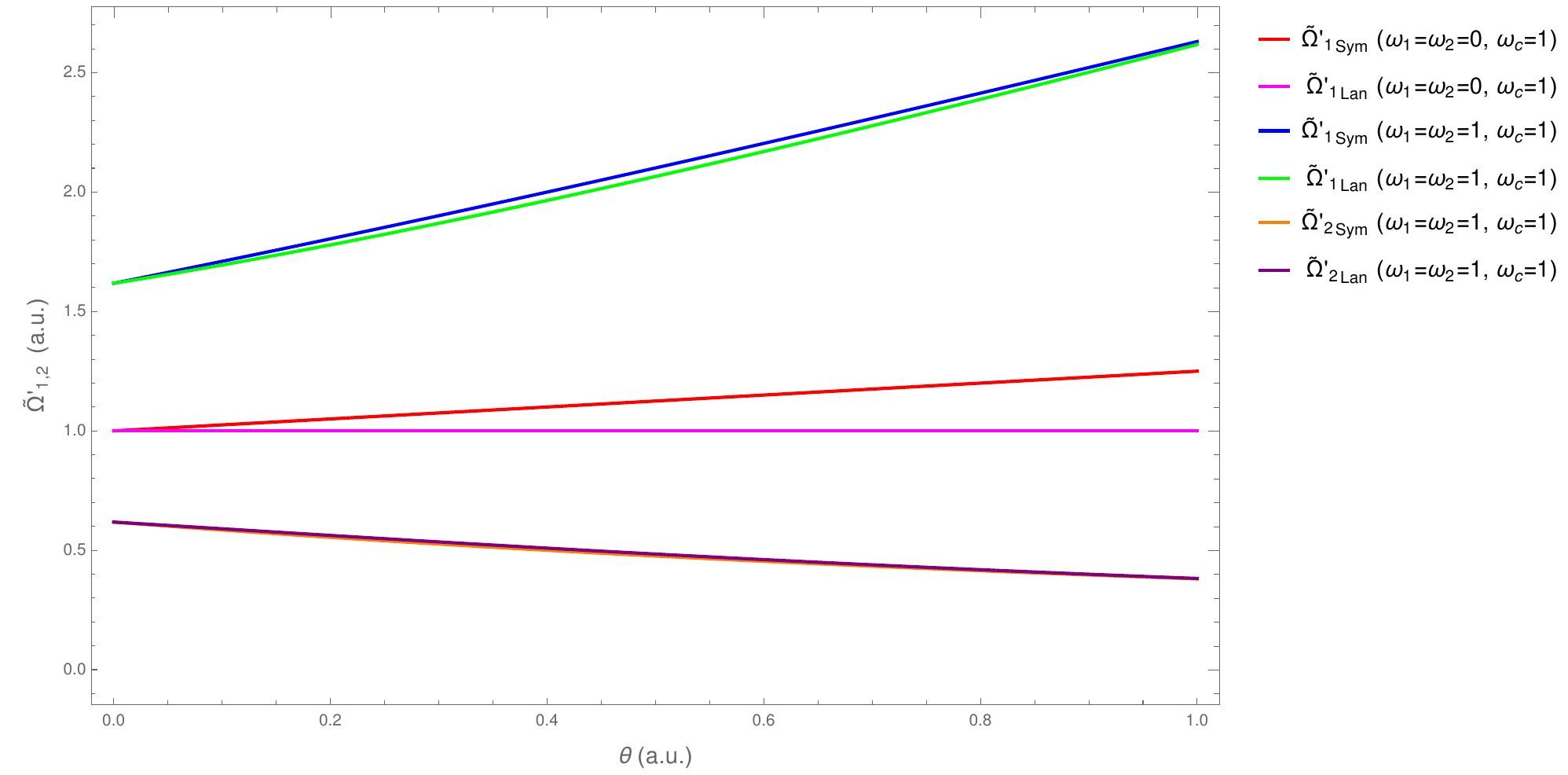}}\vspace{0cm}
\caption{In the case of naive minimal prescription, both of the eigenfrequencies $\tilde{\Omega}'_{1,2}$ as a function of $\vartheta$ in arbitrary unit (a.u.)  differ for the symmetric (Sym) and Landau (Lan) gauges  in case of pure Landau ($\omega_{1}=\omega_{2}=0$) and isotropic harmonic potential ($\omega_{1}=\omega_{2}=1$ a.u.). Here also, $\hbar=1$, $m=1$ and $\omega_{c}=1$ in arbitrary unit.}
\label{ns1fig}
\end{figure}

\begin{figure}[h!]
\centerline{\includegraphics[width=14cm]{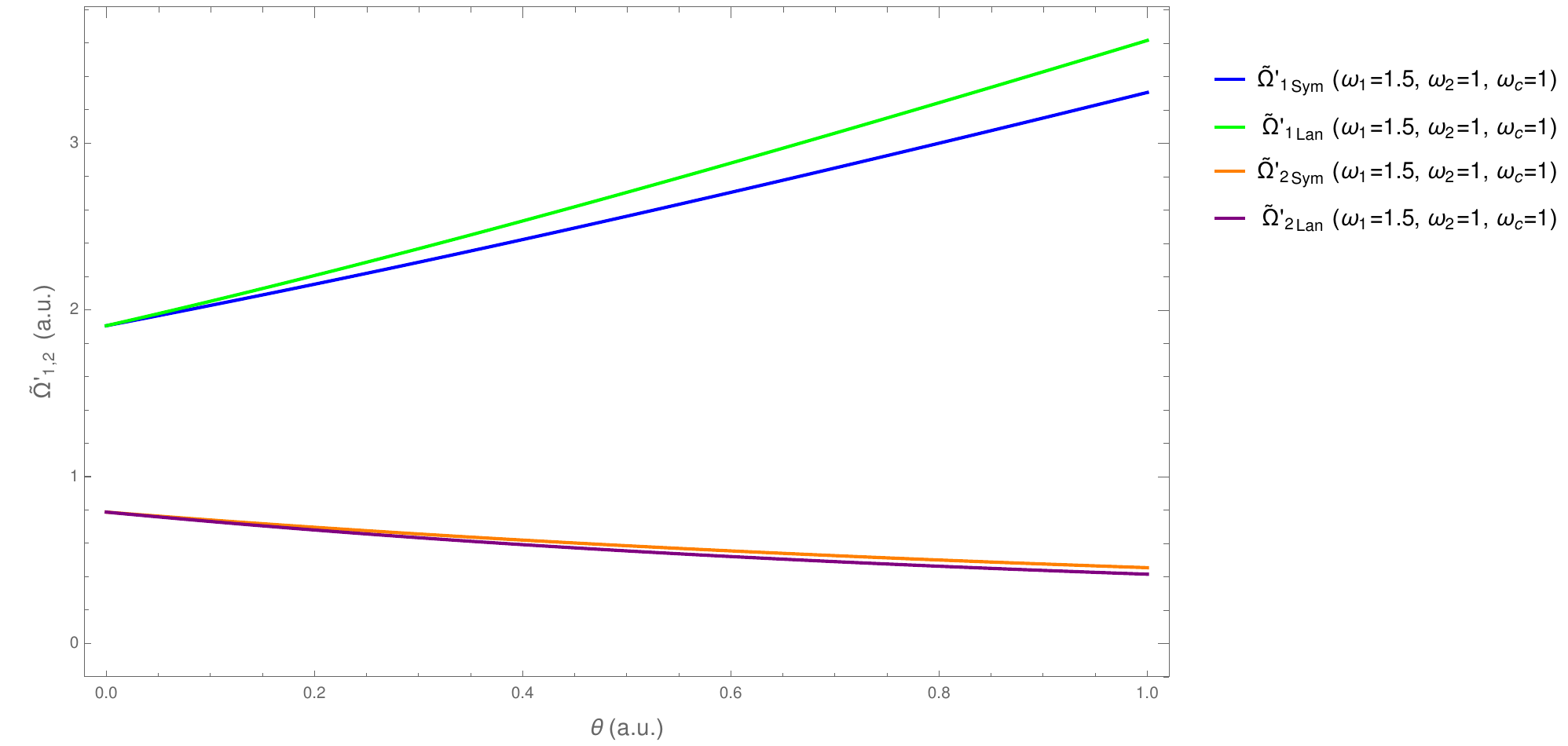}}\vspace{0cm}
\caption{In the case of naive minimal prescription, both of the eigenfrequencies $\tilde{\Omega}'_{1,2}$ as a function of $\vartheta$ in arbitrary unit (a.u.)  differ for the symmetric (Sym) and Landau (Lan) gauges  in case of anisotropic harmonic potential ($\omega_{1}=1.5, \omega_{2}=1$ a.u.). Comparing Fig. \ref{ns1fig} we can see that the anisotropy in harmonic potential increases the difference between two eigenfrequencies in two gauges. Here also, $\hbar=1$, $m=1$ and $\omega_{c}=1$ in arbitrary unit.}
\label{ns2fig}
\end{figure}

As pointed out in section \ref{sec:group-theory}, the naive minimal prescription (NMP) in noncommutative Quantum Mechanical problem in the presence of a constant magnetic field fails, and results in gauge-dependency of the eigenfrequencies given in \eqref{eq:egfreq} as shown in Fig. \ref{ns1fig} and \ref{ns2fig}. In this section, we explicitly show how such gauge dependency of the eigenfrequencies arises in this context. Naive minimal prescription in such noncommutative setup is embraced in \cite{ncanistropicoscltr, gangopadhyay, dulat}.

In the case of naive minimal prescription, the kinematical momentum of NC hamiltonian in \eqref{eq:NC-Hamiltonian},
\begin{equation}\nonumber
H^{NC}=\frac{1}{2m}\big(\hat{\Pi}_x^2+\hat{\Pi}_y^2\big)+\frac{1}{2}m\big(\omega_1^2\hat{X}^2+\omega_2^2\hat{Y}^2\big).
\end{equation}
are replaced with,
\begin{equation}\label{naive momenta}
\begin{aligned}
\hat{\Pi}_{x}=\hat{p}_x-\hat{A}_{x}; \quad \hat{\Pi}_{y}=\hat{p}_{y}-\hat{A}_{y},\,\,\,(e=1,\,c=1)
\end{aligned}
\end{equation}
where, $\hat{\mathbf{A}}= \left(\hat{A}_x(\hat{X},\hat{Y}), \hat{A}_y(\hat{X},\hat{Y}\right)$ is taken as the gauge potential which is a function of NC coordinates $(\hat{X},\hat{Y})$. As mentioned before, $\hat{p}_{x}, \hat{p}_{y}$ are the commuting quantum mechanical momentum operators, i.e. $[\hat{p}_{x},\hat{p}_{y}]=0$.

Subsequently, the map, usually known as generalized Bopp shift or Seiberg-Witten map in the literature \cite{gouba}, is used to go from NC Hamiltonian \eqref{eq:NC-Hamiltonian} to the Hamiltonian, $H'$ in terms of commutative position and momentum operators,
\begin{equation}\label{eq:23}
\begin{aligned}
\hat{X}&=\hat{x}-\frac{\vartheta}{2\hbar}\hat{p}_y; \quad\hat{Y}=\hat{y}+\frac{\vartheta}{2\hbar}\hat{p}_x,\\
\hat{P}_x&=\hat{p}_x; \quad \hat{P}_y=\hat{p}_y.
\end{aligned}
\end{equation}

The parameters (here denoted as primed to distinguish them from parameters in group theoretic construction sec \ref{group-th cal}) of the Hamiltonian, $H'$,
\begin{equation}\label{eq:ns-ham}
\begin{aligned}
H'=\frac{1}{2M'_1}\hat{p}_{x}^2+\frac{1}{2M'_2}\hat{p}_{y}^2+\frac{1}{2}M'_1\Omega^{'2}_1\hat{x}^2+\frac{1}{2}M'_2\Omega^{'2}_2\hat{y}^2-l'_1\hat{x}\hat{p}_{y}+l'_2\hat{y}\hat{p}_{x},
\end{aligned}
\end{equation}
for the symmetric gauge which is taken as, $\hat{\mathbf{A}}=\left(-\frac{B}{2}\hat{Y},\,\frac{B}{2}\hat{X}\right)$, the parameters of \eqref{eq:ns-ham} are given as,
\begin{equation}\label{eq:26}
\begin{aligned}
M'_{1_{\text{Sym}}}&=\frac{m}{1+\frac{m\omega_c\vartheta}{2\hbar}+\frac{m^2\vartheta^2}{4\hbar^2}\big(\omega_2^2+\frac{\omega_c^2}{4}\big)},\,\,\,M'_{2_{\text{sym}}}=\frac{m}{1+\frac{m\omega_c\vartheta}{2\hbar}+\frac{m^2\vartheta^2}{4\hbar^2}\big(\omega_1^2+\frac{\omega_c^2}{4}\big)},\\
\Omega^{'2}_{1_{\text{Sym}}}&=\frac{m}{M'_{1_{\text{Sym}}}}\big(\omega_1^2+\frac{\omega_c^2}{4} \big),\,\,\,\Omega^{'2}_{2_{\text{Sym}}}=\frac{m}{M'_{2_{\text{sym}}}}\big(\omega_2^2+\frac{\omega_c^2}{4} \big),\\
l'_{1_{\text{Sym}}}&=\frac{1}{2}\bigg\lbrace\omega_c\big(1+\frac{m\omega_c\vartheta}{4\hbar} \big)+\frac{m\omega_1^2\vartheta}{\hbar} \bigg\rbrace,\,\,\,l'_{2_{\text{Sym}}}=\frac{1}{2}\bigg\lbrace\omega_c\big(1+\frac{m\omega_c\vartheta}{4\hbar} \big)+\frac{m\omega_2^2\vartheta}{\hbar} \bigg\rbrace.
\end{aligned} 
\end{equation}

On the other hand, for the Landau gauge, $\hat{\mathbf{A}}=\left(-B\hat{Y},\, 0\right)$, the parameters of \eqref{eq:ns-ham} are,
\begin{equation}\label{eq:27}
\begin{aligned}
M'_{1_{\text{Lan}}}&=\frac{m}{1+\frac{m\omega_c\vartheta}{\hbar}+\frac{m^2\vartheta^2}{4\hbar^2}\big(\omega_2^2+\omega_c^2\big)},\,\,\,M'_{2_{\text{Lan}}}=\frac{m}{1+\frac{m^2\vartheta^2}{4\hbar^2}\omega_1^2},\\
\Omega^{'2}_{1_{\text{Lan}}}&=\frac{m}{M'_{1_{\text{Lan}}}}\omega_1^2,\,\,\,\Omega^{'2}_{2_{\text{Lan}}}=\frac{m}{M'_{2_{\text{Lan}}}}(\omega_2^2+\omega_c^2),\\
l'_{1_{\text{Lan}}}&=\frac{m\omega_1^2\vartheta}{2\hbar},\,\,\,l'_{2_{\text{Lan}}}=\omega_c+\frac{m\omega_c^2\vartheta}{2\hbar}+\frac{m\omega_2^2\vartheta}{2\hbar}.
\end{aligned} 
\end{equation}

From Fig. \ref{ns1fig} and \ref{ns2fig}, we can see the differences in the eigenfrequencies, $\tilde{\Omega}'_{1,2}$ calculated using the naive minimal prescription, which clearly signals their gauge-dependency and therefore such prescription is inconsistent in the context of noncommutative quantum mechanics. In this prescription, the commutation relation between observables that governs the dynamics of the quantum system also turns out to be gauge dependent, and as a consequence gauge dependency of the eigenfrequencies emerges.  In the symmetric gauge, the commutation relation between the kinematical momenta given in \eqref{naive momenta} becomes
\begin{equation}
[\Pi_{x},\Pi_{y}]=i\hbar B\left(1+\frac{\vartheta}{2\hbar}\right)\mathbb{I},
\nonumber
\end{equation}
whereas, in the Landau gauge it becomes,
\begin{equation}
[\Pi_{x},\Pi_{y}]=i\hbar B\mathbb{I}.
\nonumber
\end{equation}
Contrary to the equations above, the commutation relations given in \eqref{commut-rel} in the group-theoretic construction do not depend on the choice of the gauge, and hence we rightly found out in section \ref{group-th cal} that the eigenfrequencies are gauge invariant quantities in that construction. Besides, the gauge non-covariance of the non-commutative Hamiltonian in the case of the naive minimal prescription was also pointed out in \cite{chaichianetal}.

\section{Noncommutative quantum Hall effect}\label{sec:Hall Hamiltonian}
The second system that we choose for demonstrating gauge independence of energy eigenvalues, when calculated using the 2-parameter family of irreducible representation, is that of a noncommutative Hall Hamiltonian. Once again we have the following set of commutation relations to be satisfied

\begin{equation}\label{eq:29}
[\hat{X}^s,\hat{Y}^s]=i\vartheta\mathbb{I},\,\,\,[\hat{\Pi}^{r,s}_x,\hat{\Pi}^{r,s}_y]=i\hbar B\mathbb{I}\,\,\,\text{and}\,\,\, [\hat{X}^s,\hat{\Pi}^{r,s}_x]=[\hat{Y}^s,\hat{\Pi}^{r,s}_y]=i\hbar\mathbb{I}.
\end{equation}

In addition to demonstrating the gauge independence of energy eigenvalues, we establish the gauge independence of Hall conductivity. Subsequently, we show that there is, once again, gauge dependency of the energy eigenvalues when minimal prescription is applied. The Hall conductivity, however, accidentally turns out to be gauge invariant.

\subsection{Gauge independence of Energy Eigenvalue and Hall conductivity}\label{gauge independence}

The noncommutative Hall Hamiltonian, i.e. the Hamiltonian of an electron in the constant magnetic field $\textbf{B}=B\hat{k}$ and electric field $\textbf{E}=E\hat{i}$ in noncommutative space is

\begin{equation}\label{eq:28}
H=\frac{(\hat{\Pi}^{r,s}_{x})^2}{2m}+\frac{(\hat{\Pi}^{r,s}_{y})^2}{2m}-E\hat{X}^s,
\end{equation}  
where the mass of the electron is $m$ and its charge is taken to be unity. Note that \eqref{eq:28} can be obtained from \eqref{eq: Hamiltonian-charged-particle} by substituting $V(\hat{X}^{s},\hat{Y}^{s})=-E\hat{X}^{s}$.

Now we introduce the {\em magnetic translation operator} as a 2-parameter family of operators, $\hat{\textbf{K}}^{r,s}$, that satisfies the following algebra:

\begin{equation}\label{eq:30}
\begin{aligned}
&[\hat{X}^s,\hat{K}^{r,s}_x]=i\hbar \mathbb{I},\\
&[\hat{X}^s,\hat{K}^{r,s}_y]=0,\\
&[\hat{Y}^s,\hat{K}^{r,s}_x]=0,\\
&[\hat{Y}^s,\hat{K}^{r,s}_y]=i\hbar \mathbb{I}.
\end{aligned}
\end{equation}
and,
\begin{equation}\label{eq:31}
[\hat{\Pi}^{r,s}_x,\hat{K}^{r,s}_x]=[\hat{\Pi}^{r,s}_x,\hat{K}^{r,s}_y]=[\hat{\Pi}^{r,s}_y,\hat{K}^{r,s}_x]=[\hat{\Pi}^{r,s}_y,\hat{K}^{r,s}_y]=0.
\end{equation}
The 2 components of the family of operators $\hat{\textbf{K}}^{r,s}$ can be written as the following linear combinations:
\begin{equation}\label{eq:32}
\begin{aligned}
&\hat{K}^{r,s}_x=a_1\hat{X}^s+a_2\hat{Y}^s+a_3\hat{\Pi}^{r,s}_x+a_4\hat{\Pi}^{r,s}_y,\\
&\hat{K}^{r,s}_y=b_1\hat{X}^s+b_2\hat{Y}^s+b_3\hat{\Pi}^{r,s}_x+b_4\hat{\Pi}^{r,s}_y.
\end{aligned}
\end{equation}
After evaluating the coefficients using \eqref{eq:29}, \eqref{eq:30}, and \eqref{eq:31}, we obtain the components of the operator $\hat{\textbf{K}}^{r,s}$ as follow
\begin{equation}\label{eq:33}
\begin{aligned}
&\hat{K}^{r,s}_x=\frac{\hbar B}{\vartheta B-\hbar}\hat{Y}^s-\frac{\hbar}{\vartheta B-\hbar}\hat{\Pi}^{r,s}_x,\\
&\hat{K}^{r,s}_y=-\frac{\hbar B}{\vartheta B-\hbar}\hat{X}^s-\frac{\hbar}{\vartheta B-\hbar}\hat{\Pi}^{r,s}_y,
\end{aligned}
\end{equation}

with 
\begin{equation}\label{magnetic-commutn-rel}
[\hat{K}^{r,s}_x,\hat{K}^{r,s}_y]=\frac{i\hbar^2B}{\vartheta B-\hbar} \mathbb{I}.
\end{equation}

At this point we want to write the quantum mechanical operators $\hat{x},\, \hat{y}, \, \hat{p}_x, \,\hat{p}_y$ explicitly, using \eqref{eq:algebra-rep-gauge-before-manipulation},
\begin{equation}\label{eq:34}
\begin{aligned}
&\hat{x}=\bigg[-\frac{1}{B}+\frac{(1-r)\vartheta s}{\hbar-\vartheta B r}\bigg]\hat{\Pi}^{r,s}_y+\bigg[\frac{1}{B}+\frac{(-1+s)\vartheta}{\hbar}+\frac{(-1+r)s \vartheta}{\hbar-\vartheta B r}\bigg]\hat{K}^{r,s}_y,\\
&\hat{y}=\bigg[\frac{Br(-1+s)\vartheta+\hbar}{B \hbar}\bigg]\hat{\Pi}^{r,s}_x+\bigg[\frac{B(r+s-rs)\vartheta-\hbar}{B\hbar}\bigg]\hat{K}^{r,s}_x,\\
&\hat{p}_x=r\hat{\Pi}^{r,s}_x+(1-r)\hat{K}^{r,s}_x\\
&\hat{p}_y=\bigg[\frac{\hbar(-1+r)}{\vartheta Br-\hbar}\bigg]\hat{\Pi}^{r,s}_y+\bigg[\frac{(B\vartheta-\hbar)r}{B\vartheta r-\hbar}\bigg]\hat{K}^{r,s}_y.
\end{aligned}
\end{equation}
From \eqref{eq:34}, we see that the following required commutation relations are satisfied.
\begin{equation}
\begin{aligned}
&[\hat{x},\hat{y}]=0,\\
&[\hat{x}_i,\hat{p}_j]=i\hbar \delta_{ij} \mathbb{I},\\
&[\hat{p}_x,\hat{p}_y]=0.
\end{aligned}
\end{equation}
Using \eqref{eq:34}, the Hamiltonian can now be written as
\begin{equation}
\begin{aligned}
H^{NC}&=\frac{(\hat{\Pi}^{r,s}_{x})^2}{2m}+\frac{(\hat{\Pi}^{r,s}_{y})^2}{2m}-E\big(\hat{x}-\frac{\vartheta s}{\hbar}\hat{p}_y \big),\\
H^{NC}&=\frac{(\hat{\Pi}^{r,s}_{x})^2}{2m}+\frac{(\hat{\Pi}^{r,s}_{y})^2}{2m}+\frac{E}{B}\hat{\Pi}^{r,s}_y-\big(\frac{E}{B}-\frac{E\vartheta}{\hbar}\big)\hat{K}^{r,s}_y.
\end{aligned}
\end{equation}
Now if we define $\hat{P}^{r,s} \equiv \hat{\Pi}^{r,s}_{x}$ and $\hat{Q}^{r,s}\equiv -\frac{\hat{\Pi}^{r,s}_{y}}{B}-\frac{mE}{B^2}\mathbb{I}$, which satisfy the commutation relation, $[\hat{Q}^{r,s},\hat{P}^{r,s}]=i\hbar \mathbb{I}$, the Hamiltonian can be written as,
\begin{equation}
H^{NC}=\bigg[\dfrac{(\hat{P}^{r,s})^2}{2m}+\dfrac{B^2}{2m}(\hat{Q}^{r,s})^2\bigg]-   \frac{mE^2}{2B^2} \mathbb{I}-\bigg(\frac{E}{B}-\frac{E\vartheta}{\hbar}\bigg)\hat{K}^{r,s}_y,
\end{equation}
where the term in the square bracket is the well known Hamiltonian of a displaced harmonic oscillator and the Hamiltonian can then be written as,
\begin{equation}
H^{NC}=\hbar\omega_c\bigg[\hat{b}^{\dagger}\hat{b}+\frac{1}{2} \mathbb{I}\bigg]-   \frac{mE^2}{2B^2} \mathbb{I}-\bigg(\frac{E}{B}-\frac{E\vartheta}{\hbar}\bigg)\hat{K}^{r,s}_y,
\end{equation}
where $\hat{b}^\dagger$ and $\hat{b}$ are the ladder operators of the harmonic oscillator Hamiltonian given by
\begin{equation}
\begin{aligned}
-\frac{\hat{\Pi}^{r,s}_{y}}{B}-\frac{mE}{B^{2}}\mathbb{I}&=\sqrt{\frac{\hbar}{2B}}(\hat{b}^{\dagger}+\hat{b}),\\
\hat{\Pi}^{r,s}_{x}&=i\sqrt{\frac{\hbar B}{2}}(\hat{b}^{\dagger}-\hat{b}).
\end{aligned}
\end{equation}

Therefore, using the energy eigenstate of 1D harmonic oscillator we get the following eigenvalue equation:
\begin{equation}
E_n=\hbar\omega_c(n+1/2)-\dfrac{mE^2}{2B^2}-\bigg(\dfrac{E}{B}-\dfrac{E\vartheta}{\hbar}\bigg)\alpha,
\end{equation}
where the eigenstate of harmonic oscillator, $\ket{n}$, is also eigenstate of $\hat{K}^{r,s}_y$ with $\hat{K}^{r,s}_y\ket{n}=\alpha\ket{n}, \, \alpha\in \mathbb{R}$.
So, we see that the eigenvalues are independent of the $r, \, s$ parameters, i.e. the energy eigenvalues are gauge invariant.\\

Now, we want to calculate the Hall conductivity and check its gauge invariance. First we calculate the current operator:
\begin{equation}
\hat{J}_x=\frac{\rho}{i\hbar}[\hat{X}^s,H^{NC}]=\frac{\rho}{m}\hat{\Pi}^{r,s}_x,
\end{equation}
and,
\begin{equation}
\begin{aligned}
\hat{J}_y&=\frac{\rho}{i\hbar}[\hat{Y}^s,H^{NC}],\\
&=\frac{\rho}{m}\hat{\Pi}^{r,s}_y+\frac{\rho E\vartheta}{\hbar} \mathbb{I},\\
\hat{J}_y&=-\frac{\rho}{m}\sqrt{\frac{\hbar B}{2}}(\hat{b}+\hat{b}^{\dagger})-\frac{\rho E}{B} \mathbb{I}+\frac{\rho E\vartheta}{\hbar} \mathbb{I},
\end{aligned}
\end{equation}
where we have used $\hat{\Pi}^{r,s}_y=-\sqrt{\frac{\hbar B}{2}}(\hat{b}+\hat{b}^{\dagger})-\frac{mE}{B} \mathbb{I}$. The expectation values of the two components of the current operator are then,
\begin{equation}
\begin{aligned}
&\bra{n}\hat{J}_x\ket{n}=0,\\
&\bra{n}\hat{J}_y\ket{n}=-\rho\bigg(\frac{E}{B}-\frac{E \vartheta}{\hbar}\bigg).
\end{aligned}
\end{equation}
Therefore the Hall conductivity, $\sigma_H$, is,
\begin{equation}\label{eq:group-th-hall-conductivity}
\sigma_H=-\frac{\rho}{B}+\frac{\rho \vartheta}{\hbar}.
\end{equation}
So, we see here that the Hall conductivity, when calculated using the 2-parameter family of self-adjoint irreducible representations of $\mathcal{U}(\G)$, is also gauge invariant.

\subsection{Naive Minimal Prescription}\label{Naive Hall}
Similar to the case of anisotropic harmonic potential in section \ref{naive-minimal}, here in this section we show that the eigenvalues of the Hall Hamiltonian also turn out to be gauge dependent when calculated using the naive minimal prescription, as has been done, for example, in \cite{Dayietal,dulat2009quantum }. \\
The Hall Hamiltonian is given by
\begin{equation}
H^\prime=\frac{(\hat{\Pi}_{x})^2}{2m}+\frac{(\hat{\Pi}_{y})^2}{2m}-E\hat{X},
\end{equation}
where the kinematical momenta and the noncommutative coordinates are again given by equation \eqref{naive momenta} and equation \eqref{eq:23}, respectively.
\begin{equation*}
\begin{aligned}
\hat{\Pi}_{x}=\hat{p}_x-\hat{A}_{x}; \quad \hat{\Pi}_{y}=\hat{p}_{y}-\hat{A}_{y},\,\,\,(e=1,\,c=1),
\end{aligned}
\end{equation*}\\
and
\begin{equation*}
\begin{aligned}
\hat{X}&=\hat{x}-\frac{\vartheta}{2\hbar}\hat{p}_y; \quad\hat{Y}=\hat{y}+\frac{\vartheta}{2\hbar}\hat{p}_x,\\
\hat{P}_x&=\hat{p}_x; \quad \hat{P}_y=\hat{p}_y.
\end{aligned}
\end{equation*}\\
First we work out the Landau gauge case, i.e. $\hat{\textbf{A}}=(0,B\hat{X})$. Hence, the kinematical momenta are of the following form, 
\begin{equation}\label{eq:35}
\begin{aligned}
&\hat{\Pi}_{x_{\text{Lan}}}=\hat{p}_x,\\
&\hat{\Pi}_{y_{\text{Lan}}}=\gamma_{\text{Lan}}\hat{p}_y-B\hat{x},
\end{aligned}
\end{equation}
where $\gamma_{\text{Lan}} \equiv 1+ \frac{\vartheta B}{2 \hbar}$.
Consequently, we have the following set of commutation relations:
\begin{equation}
\begin{aligned}
&[\hat{x},\hat{\Pi}_{x_{\text{Lan}}}]=i\hbar \mathbb{I},\,\,\,[\hat{y},\hat{\Pi}_{y_{\text{Lan}}}]=i\hbar \gamma_{\text{Lan}} \mathbb{I},\\
&[\hat{\Pi}_{x_{\text{Lan}}},\hat{\Pi}_{y_{\text{Lan}}}]=i\hbar B \mathbb{I},\,\,\,[\hat{x},\hat{\Pi}_{y_{\text{Lan}}}]=[\hat{y},\hat{\Pi}_{x_{\text{Lan}}}]=0.
\end{aligned}
\end{equation}
We again introduce the magnetic translation operator for the Landau gauge, $\hat{\textbf{K}}_\text{Lan}$ whose components now obey the following set of commutation relations:
\begin{equation}
\begin{aligned}
&[\hat{x},\hat{K}_{x_\text{Lan}}]=i\hbar \mathbb{I},\,\,\, [\hat{y},\hat{K}_{y_\text{Lan}}]=i\hbar \mathbb{I},\\
&[\hat{x},\hat{K}_{y_\text{Lan}}]=0,\,\,\, [\hat{y},\hat{K}_{x_\text{Lan}}]=0.
\end{aligned}
\end{equation}
and
\begin{equation}
[\hat{\Pi}_{x_\text{Lan}},\hat{K}_{x_\text{Lan}}]=[\hat{\Pi}_{x_\text{Lan}},\hat{K}_{y_\text{Lan}}]=[\hat{\Pi}_{y_\text{Lan}},\hat{K}_{x_\text{Lan}}]=[\hat{\Pi}_{y_\text{Lan}},\hat{K}_{y_\text{Lan}}]=0.
\end{equation}

Using the above commutation relations $\hat{\textbf{K}}_\text{Lan}$ can be written as:
\begin{equation}\label{eq:36}
\begin{aligned}
&\hat{K}_{x_{\text{Lan}}}=\hat{\Pi}_{x_{\text{Lan}}}-\frac{B}{\gamma_{\text{Lan}}}\hat{y},\\
&\hat{K}_{y_{\text{Lan}}}= \frac{1}{\gamma_{\text{Lan}}}\hat{\Pi}_{y_{\text{Lan}}}+\frac{ B}{\gamma_{\text{Lan}}}\hat{x},
\end{aligned}
\end{equation} 
with the commutation relation:
\begin{equation}
[\hat{K}_{x_{\text{Lan}}},\hat{K}_{y_{\text{Lan}}}]=-\frac{i\hbar B}{\gamma_{\text{Lan}}} \mathbb{I}.
\end{equation}
Using \eqref{eq:35} with \eqref{eq:36}, the commutative coordinates then can be written down as
\begin{equation}
\begin{aligned}
&\hat{x}=\frac{\gamma_{\text{Lan}}}{ B}\hat{K}_{y_{\text{Lan}}}-\frac{1}{B}\hat{\Pi}_{y_{\text{Lan}}},\\
&\hat{y}=\frac{\gamma_{\text{Lan}}}{ B}\hat{\Pi}_{x_{\text{Lan}}}-\frac{\gamma_{\text{Lan}}}{ B}\hat{K}_{x_{\text{Lan}}},\\
&\hat{p}_x=\hat{K}_{x_{\text{Lan}}}+\frac{B}{\gamma_{\text{Lan}}}\hat{y},\\
&\hat{p}_y=\hat{K}_{y_{\text{Lan}}}.
\end{aligned}
\end{equation}
The Hall Hamiltonian in Landau gauge given by
\begin{equation}
H^\prime_{\text{Lan}}=\frac{(\hat{\Pi}_{x_{\text{Lan}}})^2}{2m}+\frac{(\hat{\Pi}_{y_{\text{Lan}}})^2}{2m}-E\bigg(\hat{x}-\frac{\vartheta}{2\hbar}\hat{p}_y\bigg),
\end{equation}
then becomes
\begin{equation}
H^\prime_{\text{Lan}}=\frac{(\hat{\Pi}_{x_{\text{Lan}}})^2}{2m}+\frac{(\hat{\Pi}_{y_{\text{Lan}}})^2}{2m}+\frac{E}{B}\hat{\Pi}_{y_{\text{Lan}}}-\frac{E}{B}\hat{K}_{y_{\text{Lan}}},
\end{equation} 
which, in turn, can be cast into the following form
\begin{equation}\label{eq:landau-ham-square}
H^\prime_{\text{Lan}}=\bigg[\frac{(\hat{\Pi}_{x_{\text{Lan}}})^2}{2m}+\frac{1}{2}m\omega_{c}^{2}\bigg(\frac{\hat{\Pi}_{y_{\text{Lan}}}}{B}+\frac{mE}{B^{2}}\mathbb{I}\bigg)^{2}\bigg]-\frac{mE^2}{2B^2} \mathbb{I}-\frac{E}{B}\hat{K}_{y_{\text{Lan}}}.
\end{equation}

The Hamiltonian piece inside the square bracket is reminiscent of that of a 1D displaced harmonic oscillator so that equation \eqref{eq:landau-ham-square} can now be recast as
\begin{equation}
H^\prime_{\text{Lan}}=\hbar\omega_c\bigg(\hat{c}^\dagger \hat{c}+\frac{1}{2}\mathbb{I}\bigg)-\frac{mE^2}{2B^2} \mathbb{I}-\frac{E}{B}\hat{K}_{y_{\text{Lan}}},
\end{equation} 
where $\hat{c}^\dagger$ and $\hat{c}$ are ladder operators such that the position and momentum of the displaced 1D harmonic oscillator in terms of the ladder operators can be read off as
\begin{equation}
\begin{aligned}
&\hat{Q}_\text{Lan}:=-\frac{\hat{\Pi}_{y_{\text{Lan}}}}{B}-\frac{mE}{B^2} \mathbb{I}=\sqrt{\frac{\hbar}{2B}}(\hat{c}^\dagger+\hat{c}),\\
&\hat{P}_\text{Lan}:=\hat{\Pi}_{x_{\text{Lan}}}=i\sqrt{\frac{\hbar B}{2}}(\hat{c}^\dagger-c).
\end{aligned}
\end{equation}
Therefore, upon using the energy eigenstates of a 1D harmonic oscillator, we obtain the following eigenvalues:
\begin{equation}\label{EL}
E^\prime_{\text{Lan}}=\hbar\omega_c\bigg(n+\frac{1}{2}\bigg)-\frac{mE^2}{2B^2}-\frac{E}{B}\alpha.
\end{equation}
It is interesting to note that in Landau gauge, the eigenvalue for a noncommutative Hall Hamiltonian is the same as that of a commutative Hall Hamiltonian.\\

Now, we proceed to calculate the Hall conductivity. The components of the current operator read
\begin{equation}
\hat{J}_{x_{\text{Lan}}}=\frac{\rho}{i\hbar}[\hat{x},H^\prime_{\text{Lan}}]=\frac{\rho}{m}\hat{\Pi}_{x_{\text{Lan}}},
\end{equation}
and
\begin{equation}
\begin{aligned}
\hat{J}_{y_{\text{Lan}}}=\frac{\rho}{i\hbar}[\hat{y},H^\prime_{\text{Lan}}]=\frac{\rho B \gamma_{\text{Lan}}}{m}\hat{Q}_{\text{Lan}}-\frac{\rho E}{B}  \mathbb{I}.
\end{aligned}
\end{equation}
They have the following expectation values:
\begin{equation}
\begin{aligned}
&\bra{n}\hat{J}_{x_{\text{Lan}}}\ket{n}=0,\\
&\bra{n}\hat{J}_{y_{\text{Lan}}}\ket{n}=-\frac{\rho E}{B}.
\end{aligned}
\end{equation}
Hence the Hall conductivity is given by
\begin{equation}\label{eq:hall-conductivity-naive-lan}
\sigma_{H_{\text{Lan}}}=-\frac{\rho}{B}.
\end{equation}

Next we repeat the calculation in symmetric gauge, i.e. $\hat{\textbf{A}}=\big(\!-\frac{B}{2}\hat{Y},\frac{B}{2}\hat{X}\big)$. The kinematical momenta in symmetric gauge read
\begin{equation}
\begin{aligned}
&\hat{\Pi}_{x_{\text{Sym}}}=\gamma_{\text{Sym}}\hat{p}_x+\frac{B}{2}\hat{y},\\
&\hat{\Pi}_{y_{\text{Sym}}}=\gamma_{\text{Sym}}\hat{p}_y-\frac{B}{2}\hat{x}.
\end{aligned}
\end{equation}

where $\gamma_{\text{Sym}}\equiv 1+\frac{\vartheta B}{4\hbar}$. Therefore, we have the following set of commutation relations:
\begin{equation}
\begin{aligned}
&[\hat{\Pi}_{x_{\text{Sym}}},\hat{\Pi}_{y_{\text{Sym}}}]=i\hbar B \gamma_{\text{Sym}}\mathbb{I},\\
&[\hat{x},\hat{\Pi}_{x_{\text{Sym}}}]=i\hbar\gamma_{\text{Sym}} \mathbb{I},\\
&[\hat{y},\hat{\Pi}_{y_{\text{Sym}}}]=i\hbar\gamma_{\text{Sym}} \mathbb{I},\\
&[\hat{x},\hat{\Pi}_{y_{\text{Sym}}}]=[\hat{y},\hat{\Pi}_{x_{\text{Sym}}}]=0.
\end{aligned}
\end{equation}
Once again we have the magnetic translation operator $\hat{\textbf{K}}_{\text{Sym}}$ in the symmetric gauge, which can be seen to have the following form in terms of its components:

\begin{equation}
\begin{aligned}
&\hat{K}_{x_{\text{Sym}}}=-\frac{ B}{\gamma_{\text{Sym}}}\hat{y}+ \frac{1}{\gamma_{\text{Sym}}} \hat{\Pi}_{x_{\text{Sym}}},\\
&\hat{K}_{y_{\text{Sym}}}=\frac{B}{\gamma_{\text{Sym}}}\hat{x}+\frac{1}{\gamma_{\text{Sym}}}\hat{\Pi}_{y_{\text{Sym}}}.
\end{aligned}
\end{equation}
They obey the following commutation relation:
\begin{equation}
[\hat{K}_{x_{\text{Sym}}},\hat{K}_{y_{\text{Sym}}}]=-\frac{i\hbar B}{\gamma_{\text{Sym}}} \mathbb{I}.
\end{equation}\\
And the commutative coordinates $\hat{x}$ and $\hat{p}_y$ can in turn be written as,
\begin{equation}
\begin{aligned}
&\hat{x}=-\frac{1}{B}\hat{\Pi}_{y_{\text{Sym}}}+\frac{\gamma_{\text{Sym}}}{B}\hat{K}_{y_{\text{Sym}}},\\
&\hat{p}_y=\frac{1}{2\gamma_{\text{Sym}}} \hat{\Pi}_{y_{\text{Sym}}}+\frac{\hat{K}_{y_{\text{Sym}}}}{2}.
\end{aligned}
\end{equation}
Then the Hamiltonian becomes
\begin{equation}
\begin{aligned}
&H^\prime_{\text{Sym}}=\frac{(\hat{\Pi}_{x_{\text{Sym}}})^2}{2m}+\frac{(\hat{\Pi}_{y_{\text{Sym}}})^2}{2m}-E\bigg(\hat{x}-\frac{\vartheta}{2\hbar}\hat{p}_y\bigg),\\
&H^\prime_{\text{Sym}}=\frac{(\hat{\Pi}_{x_{\text{Sym}}})^2}{2m}+\frac{(\hat{\Pi}^\prime_{y_{\text{Sym}}})^2}{2m}-\frac{m t^2}{2}\mathbb{I}-\frac{E}{B}\hat{K}_{y_{\text{Sym}}},
\end{aligned}
\end{equation}
where
\begin{equation}
\begin{aligned}
&t\equiv\frac{E}{B}+\frac{E\vartheta}{4\hbar\gamma_{\text{Sym}}}\;\;\hbox{and}\\
&\hat{\Pi}^\prime_{y_{\text{Sym}}}\equiv\hat{\Pi}_{y_{\text{Sym}}}+mt\mathbb{I}.
\end{aligned}
\end{equation}
Now if we define, $\hat{P}_\text{Sym}\equiv\hat{\Pi}^\prime_{y_{\text{Sym}}}$ and $\hat{Q}_\text{Sym}\equiv\frac{\hat{\Pi}_{x_{\text{Sym}}}}{B\gamma_{\text{Sym}}}$, then the underlying Hamiltonian can be recast in the following form:
\begin{equation}
\begin{aligned}
&H^\prime_{\text{Sym}}=\bigg[\frac{(\hat{P}_{\text{Sym}})^2}{2m}+\frac{1}{2}m\tilde{\omega}_c^2\hat{Q}^2_\text{Sym}\bigg]-\frac{m t^2}{2}\mathbb{I}-\frac{E}{B}\hat{K}_{y_{\text{Sym}}},
\end{aligned}
\end{equation}
where $\tilde{\omega}_c\equiv\frac{B \gamma_{\text{Sym}}}{m}$. The position and momentum operators of the harmonic oscillator Hamiltonian in the square bracket in terms of the underlying ladder operators are as follow
\begin{equation}
\begin{aligned}
&\hat{Q}_{\text{Sym}}=\sqrt{\frac{\hbar}{2B\gamma_{\text{Sym}}}}(\hat{a}^{\dagger}+\hat{a}),\\
&\hat{P}_{\text{Sym}}=i\sqrt{\frac{\hbar B\gamma_{\text{Sym}}}{2}}(\hat{a}^{\dagger}-\hat{a}),
\end{aligned}
\end{equation}  
which allows us to write the total Hamiltonian as
\begin{equation}
\begin{aligned}
&H^\prime_{\text{Sym}}=\hbar \tilde{\omega}_c\bigg(\hat{a}^\dagger\hat{a}+\frac{1}{2}\bigg)-\frac{m t^2}{2}\mathbb{I}-\frac{E}{B}\hat{K}_{y_{\text{Sym}}}.
\end{aligned}
\end{equation}

Using the eigenstates of a 1-D harmonic oscillator, one obtains the following eigenvalues
 
\begin{equation}
E^\prime_{\text{Sym}}=\hbar\omega_c\gamma_{\text{Sym}}\bigg(n+\frac{1}{2}\bigg)-\frac{m E^2}{2B^2}\bigg( 1+\frac{\vartheta B}{4 \hbar \gamma_{\text{Sym}} }\bigg )^2-\frac{E}{B}\alpha.
\end{equation}
These eigenvalues turn out to be different from $E^\prime_{\text{Lan}}$ in \eqref{EL}. Therefore, we find that the eigenvalues are indeed gauge dependent.\\

Finally, we calculate the Hall conductivity in symmetric gauge. The components of the current operator are given by
\begin{equation}
\begin{aligned}
\hat{J}_{x_{\text{Sym}}}&=\frac{\rho}{i\hbar}[\hat{x},H^\prime_{\text{Sym}}]
=\frac{\rho \gamma_{\text{Sym}}}{m}\hat{\Pi}_{x_\text{Sym}},\\\\
\hat{J}_{y_{\text{Sym}}}&=\frac{\rho}{i\hbar}[\hat{y},H^\prime_{\text{Sym}}]
=\frac{\rho \gamma_{\text{Sym}}}{m}\hat{P}_{\text{Sym}}-\frac{\rho E}{B}\mathbb{I},
\end{aligned}
\end{equation}

which have the following expectation values
\begin{equation}
\begin{aligned}
&\bra{n}\hat{J}_{x_{\text{Sym}}}\ket{n}=0,\\
&\bra{n}\hat{J}_{y_{\text{Sym}}}\ket{n}=-\frac{\rho E}{B}.
\end{aligned}
\end{equation}
Therefore, we observe that despite the gauge dependency of the underlying components of the current operator, the gauge dependent terms drop out while calculating the Hall conductivity and it turns out to be the same for symmetric gauge as that of Landau gauge:
\begin{equation}\label{eq:hall-conductivity-naive-sym}
\sigma_{H_{\text{Sym}}}=-\frac{\rho}{B}.
\end{equation}

\section{Conclusion}\label{sec:conclusion}
In this paper, we have shown that the 2-parameter family of gauge equivalent irreducible self-adjoint representations \eqref{eq:algebra-rep-gauge-before-manipulation} of $\G$, obtained in \cite{nctori}, indeed yields gauge invariant energy spectra for an electron in a noncommutative plane subjected to a vertical constant magnetic field under the influence of a general class of polynomial potentials. We have surveyed the literature finding no uniform and consistent gauge prescription in the noncommutative setting as opposed to the standard quantum mechanical Landau problem. We also notice that the expressions for kinematical momenta $\hat{\Pi}^{r,s}_{x}$ and $\hat{\Pi}^{r,s}_{y}$ in \eqref{eq:algebra-rep-gauge-before-manipulation} reduce to the ones obeying familiar minimal prescription for quantum mechanics in the presence of a constant magnetic field  when the spatial noncommutativity parameter $\vartheta\rightarrow 0$. 

We have shown in the paper that minimal coupling prescription fails to achieve gauge invariant energy eigenvalues for the cases of anisotropic harmonic potential and the Hall potential. We diagonalize the Hall Hamiltonian by introducing the 2-parameter family of magnetic translation operators whose components satisfy the commutation relation \eqref{magnetic-commutn-rel} that is in exact agreement with (eq (26) on p. 271 of \cite{nairetal}). The gauge invariant Hall conductivity (see \eqref{eq:group-th-hall-conductivity}) that one obtains by using self-adjoint representations of $\mathcal{U}(\G)$ differ from the one (\eqref{eq:hall-conductivity-naive-lan} and \eqref{eq:hall-conductivity-naive-sym}) obtained using naive minimal coupling prescription. The noncommutative Hall conductivity obtained in the group theoretic method contains the desired $\vartheta$-correction while the one obtained using minimal coupling prescription is devoid of any $\vartheta$-correction.

In differential geometric language, the vector potential $A=A_{1}dx+A_{2}dy$ is simply a 1-form on the pertinent configuration space. Here $A_{i}\mbox{'}s$ with $i=1,2$, are smooth functions of the coordinates. Then the exterior derivative $dA$ of the gauge potential is interpreted as the applied magnetic field $B$. Quantum mechanics in the presence of a constant external magnetic field can be described in this geometric language as the underlying position operators $\hat{x}, \hat{y}$ are simply multiplication operators. But in the noncommutative setting, the $A_{i}\mbox{'}s$ as defined in (p.10 of \cite{nctori}) are no longer smooth functions of $x$ and $y$, rather are operators on $L^{2}(\mathbb{R}^{2},dxdy)$, so that the gauge potentials are now operator valued 1-forms. Exterior derivatives of operator valued 1-forms were considered in (p. 321, \cite{alvarezetal}). But we look forward to approaching the problem using Conne's construction \cite{connesbook,connespaper} by finding an appropriate spectral triple. The underlying Dirac operator then will take care of the interpretation of the exterior derivative of the operator valued 1-forms. We wish to carry out these geometric studies pertinent to the definition of vector potential suggested in \cite{nctori} in a future publication.

\subsubsection*{Acknowledgement}
We would like to thank M.M. Sheikh-Jabbari, Arshad Momen, Rakibur Rahman and Mir Mehedi Faruk for useful discussions. T.A.C would also like to thank The Abdus Salam International Centre for Theoretical Physics, where part of this work had been done, for the hospitality and the support.

\end{document}